\documentclass[a4paper,12pt]{article}
\pdfoutput=1

\usepackage{amsmath,amsfonts,amssymb,cite}
\usepackage{graphicx}
\usepackage{braket}

\def\unit{\relax{\rm 1\kern-.26em I}}

\usepackage{verbatim,setspace}
\usepackage{xparse}
\usepackage{tikz}
\usetikzlibrary{calc}
\DeclareDocumentCommand{\hcancel}{mO{0pt}O{0pt}O{0pt}O{0pt}}{%
    \tikz[baseline=(tocancel.base)]{
        \node[inner sep=0pt,outer sep=0pt] (tocancel) {#1};
        \draw[black] ($(tocancel.south west)+(#2,#3)$) -- ($(tocancel.north east)+(#4,#5)$);
    }
}

\def\fdir{figs/}

\let\oldincludegraphics\includegraphics
\renewcommand{\includegraphics}[2][]{\oldincludegraphics[#1]{\fdir #2}}

\def\twomat[#1,#2][#3,#4]{\left( \begin{array}{cc} #1 & #2 \\ #3 & #4 \end{array} \right)}\usepackage{hyperref}
\def\threemat[#1,#2,#3][#4,#5,#6][#7,#8,#9]{\left( \begin{array}{ccc} #1 & #2 & #3\\ #4 & #5 & #6 \\ #7 & #8 & #9 \end{array} \right)}
\def\twovec[#1,#2]{\left( \begin{array}{c} #1  \\ #2 \end{array} \right)}
\def\bra{\langle}
\def\ket{\rangle}
\def\C{\mathcal}

\usepackage{amsmath}
\usepackage{amssymb}
\usepackage{amsfonts}
\usepackage{epsfig}
\usepackage{soul}

\newcommand{\cH}{{\cal H}}

\newcommand{\cL}{{\cal L}}

\newcommand{\cO}{{\cal O}}

\newcommand{\ov}{\overline}

\newcommand{\ra}{\rightarrow}
\newcommand{\be}{\begin{equation}}
\newcommand{\ee}{\end{equation}}
\newcommand{\bea}{\begin{eqnarray}}
\newcommand{\eea}{\end{eqnarray}}

\DeclareMathSymbol{\mg}{\mathrel}{symbols}{"1D}

\newcommand{\MET}{E\llap{/\kern1.5pt}_T}
\newcommand{\nn}{\nonumber}

\numberwithin{equation}{section}

\newcounter{oldcounter}
\addtocounter{equation}{1}
\begin{document}

 \begin{flushright}
{CPHT099.1213}
\end{flushright}

\thispagestyle{empty}

\begin{center}
\begin{spacing}{2}
{\LARGE {\bf Flavour models with Dirac and fake gluinos}}
\end{spacing}
\vspace{0.5cm}
{\bf Emilian Dudas$^{a,}$\footnote{emilian.dudas@cpht.polytechnique.fr},
Mark Goodsell$^{b,}$\footnote{mark.goodsell@lpthe.jussieu.fr}, Lucien Heurtier$^{a,}$\footnote{heurtier@cpht.polytechnique.fr}\\ and  Pantelis Tziveloglou$^{c,d,}$\footnote{pantelis.tziveloglou@vub.ac.be}}
\end{center}

\noindent {\small $^a$Centre de Physique Th\'eorique, \'Ecole Polytechnique, CNRS, 91128 Palaiseau, France\\
$^b$Laboratoire de Physique Th\'eorique et Hautes Energies, CNRS, UPMC Universit\'e Paris VI, Boite 126, 4 Place Jussieu, 75252 Paris cedex 05, France \\
$^c$Theoretische Natuurkunde and IIHE, Vrije Universiteit Brussel, Pleinlaan 2,\\ B-1050 Brussels, Belgium \\
$^d$International Solvay Institutes, Brussels, Belgium
}

\vspace{0.3cm}


\begin{abstract}
\noindent In the context of supersymmetric models where the gauginos may have both Majorana and Dirac masses
we investigate the general constraints from flavour-changing processes on the scalar mass matrices. One finds
that the chirality-flip suppression of flavour-changing effects usually invoked in the
pure Dirac case holds in the mass insertion approximation but not in the general case, and fails 
in particular for inverted hierarchy models. We quantify the constraints in several
flavour models which correlate fermion and scalar superpartner masses. We also discuss the
limit of very large Majorana gaugino masses compared to the chiral adjoint and Dirac masses, where the 
remaining light eigenstate is the ``fake'' gaugino, including the consequences 
of suppressed couplings to quarks beyond flavour constraints.
\end{abstract}

\newpage

\tableofcontents

\setcounter{page}{1}

\newpage

\section{Introduction}
\label{SEC:Introduction}

Supersymmetric extensions of the Standard Model are arguably still the 
most plausible ways to deal with the various mysteries of the Standard Model.
The absence of a new physics signature at LHC for the time being suggests, however,
that we should seriously (re)consider non-minimal extensions compared to the minimal supersymmetric
extension (MSSM) in all its various forms. Furthermore, it has been known since the early
days of low-energy supersymmetry that flavour-changing processes set severe constraints on the flavour structure of the superpartner spectrum in the MSSM. For example, the simplest models
based on a single abelian flavoured gauge group, although providing an approximate alignment mechanism for scalar mass matrices, still require scalar partners heavier than 
at least $100$ TeV.  Both collider and flavour constraints encourage us to search for non-minimal extensions with suppressed collider
bounds and flavour-changing transitions. Supersymmetric extensions with a Dirac gaugino
sector \cite{fayet,Polchinski:1982an,Hall:1990hq,Nelson:2002ca,fnw,Antoniadis:2005em,Antoniadis:2006uj,kpw,Amigo:2008rc,Plehn:2008ae,Benakli:2008pg,Belanger:2009wf,Benakli:2009mk,Choi:2009ue,Benakli:2010gi,Choi:2010gc,Carpenter:2010as,Kribs:2010md,Abel:2011dc,Davies:2011mp,Benakli:2011kz,Kalinowski:2011zz,Frugiuele:2011mh,ItoyamaMaru,Rehermann:2011ax,Bertuzzo:2012su,Davies:2012vu,Argurio:2012cd,Fok:2012fb,Argurio:2012bi,Frugiuele:2012pe,Frugiuele:2012kp,Benakli:2012cy,Chakraborty:2013gea,Csaki:2013fla,MDGSSM} enter precisely into this category. 

Originally motivated by the preserved R-symmetry, which allows simpler supersymmetry breaking sectors \cite{fayet,Polchinski:1982an}, and the possible connection with extra dimensions and $N=2$ supersymmetry \cite{Antoniadis:2005em}, it was subsequently noticed that Dirac gaugino masses have many phenomenological advantages over their Majorana counterparts. For example, the Dirac mass is \emph{supersoft} \cite{Jack:1999ud,Jack:1999fa,fnw,Goodsell:2012fm}, which naturally allows somewhat heavy 
gluinos compared to the squarks \cite{Brust:2011tb,Papucci:2011wy,Arvanitaki:2013yja}. Furthermore, it was argued later on that in this case flavour-changing neutral current (FCNC) transitions are suppressed due to protection from the underlying R-symmetry that lead to a chirality flip suppression \cite{kpw}. It was also proved that the collider signatures of superpartner production are suppressed compared to the MSSM case due to the heaviness of the Dirac gluino and the absence of several squark decay channels \cite{Heikinheimo:2011fk,kribsmartin,Kribs:2013eua,Kribs:2013oda}. The main goal of this paper is to understand the most general bounds  from flavour physics when we allow Dirac gaugino masses in addition to Majorana masses. 

We begin the paper in section \ref{SEC:Expressions} by giving the general expressions for the meson mixing ($\Delta F=2$, i.e. a change of two units of flavour) FCNC processes in models with both Dirac and Majorana gluino masses. We also introduce the notation used in the remainder of the paper. 

In much of the literature where flavour constraints are discussed, in an attempt to provide relatively model-independent bounds, scalar mass matrices are treated in the so-called mass insertion approximation, in which scalars are almost degenerate with small off-diagonal entries. Indeed, where flavour constraints in Dirac gaugino models have been considered, the mass insertion approximation was also used \cite{kpw,Fok:2012me}. Hence we first provide an updated discussion of this case in section \ref{SEC:MassInsertion}, with in addition bounds for differing ratios of Dirac and Majorana gluino masses, with no restrictions provided by the R-symmetry.

However, in particular in light of bounds on superpartner masses, the mass insertion approximation is actually rather difficult to realise in any flavour model. We are therefore led to consider general flavour models/scenarios which go beyond this approximation in section \ref{SEC:Models}. An important result is that, surprisingly, we find that the dramatic chirality-flip suppression of \cite{kpw}  is at work only in a small number of cases, whereas in the general case the suppression is much milder and in certain cases the Majorana case is \emph{less} constrained.  
Our main working assumption is that the flavour symmetry explaining the fermion masses and mixings governs simultaneously the superpartner spectrum. We find that the simplest single $U(1)$ flavour models do still need heavy scalars. For the case of two $U(1)$'s we find the unusual
feature that, in some regions of parameter space, Dirac models are more constrained than
their Majorana counterparts, due to cancellations occurring in the latter case. We also investigate the inverted hierarchy case and one example of nonabelian flavour symmetries, discuss the K (and B meson constraints in appendix \ref{APP:Bs}) and compare them with their MSSM counterpart models. 
  
As a refreshing aside, in section \ref{SEC:FakeGaugino} we consider also the unusual case where the lightest adjoint fermions couple in a suppressed way to the quarks, due to their very little gaugino component. This happens when the Majorana gaugino mass is much bigger than the Dirac and
the adjoint fermion masses. This can occur for relatively light squarks and gluinos or for intermediate scale values. In both cases light adjoint fermions have suppressed couplings to quarks, a case we refer to as ``fake gluino''\footnote{We acknowledge K.~Benakli and P.~Slavich for suggesting the name during collaboration on a related work \cite{DGSplit}.}. The first case can lead to the unusual feature
of experimentally accessible squarks, but long-lived (fake) gluinos. The intermediate scale case is  interesting from the viewpoint of gauge coupling unification. In this case, radiative corrections lead to heavy scalars and therefore the scenario is similar in spirit to split supersymmetry\cite{SplitSUSY}, but with suppressed ``fake gluino/gaugino'' couplings to quarks and to higgs/higgsinos. Since the radiative stability of this scenario requires some particular high-energy symmetries, it has specific features distinguishing it from standard split supersymmetry and other related scenarios \cite{Carena:2004ha,Antoniadis:2005em,Unwin:2012fj} which we shall discuss.

Finally, as a note to the concerned reader, in this paper we largely only discuss $\Delta F=2$ constraints arising from box diagrams involving gluinos. In principle, there are also diagrams that contribute at two loops from processes involving the octet scalar partners of the Dirac gluino, which were discussed in \cite{Plehn:2008ae} and shown to be small; similarly we do not include subdominant contributions to the box diagrams coming from electroweak gauginos/higgsinos because they do not add qualitatively to the discussion. In addition, there are also constraints coming from $\Delta F = 1$ processes such as $b \rightarrow s \gamma$, $\mu \rightarrow e \gamma$ and electric dipole moments. These have been discussed in the context of the MRSSM and the mass insertion approximation \cite{kpw,Fok:2012me}. However, with the exception of $b\rightarrow s \gamma$ these are all dependent on the Higgs structure of the theory, and not only on the squark/quark mass matrices, since the Dirac gaugino paradigm allows many possible Higgs sectors \cite{fnw,Nelson:2002ca,kpw,Benakli:2011kz,Benakli:2012cy,MDGSSM}. For example, if we insist that the model preserves an exact R-symmetry, then these processes are suppressed so much as to be negligible; but they become relevant if we allow the Higgs sector to break R-symmetry \cite{Benakli:2012cy}. Thus it is not possible to describe bounds on these in a model-independent way, and we refrain from attempting to do so.  For the case of $b \rightarrow s \gamma$, the constraints are generically weaker than the $\Delta F =2$ case, and moreover the expressions are the same in both the Majorana and Dirac cases, since they do not involve a chirality flip; they are thus irrelevant for this paper.

\section{Neutral meson mixing in supersymmetry with Dirac gauginos}
\label{SEC:Expressions}

In recent years, very precise measurements of observables in flavour violation processes have been made \cite{pdg} while the Standard Model contribution to some of these processes is now being known with reasonable accuracy \cite{utfit}. This results in very strong restrictions on the flavour structure of theories beyond the SM.

Some of the strongest constraints arise from neutral meson mixing systems, in particular the neutral $K$-, $B_d$-, $B_s$- and $D$- meson systems \cite{nirperez}. An exact theoretical computation of these processes is particularly difficult due to unresolved non-perturbative, strong-interaction effects. The general strategy is to compute the amplitude between the valence quarks in the full perturbative theory, then match the amplitude to an effective theory of four-fermion contact interactions. Contact with neutral meson mixing is achieved by estimating the matrix elements between initial and final states, typically by use of PCAC \cite{pcac} and lattice QCD techniques.

\subsection{Effective Hamiltonian}

Within the context of MSSM, the dominant contribution to neutral meson mixing comes from gluino-squark box diagrams (see e.g. figure \ref{boxdiagrams} for the Kaon system). In the following, we expand the standard computation (see app. \ref{MSSMbox}) to include both Majorana and Dirac gluino masses. In particular
\bea\label{Ldirac}
\cL&\supset&-{1\over 2}\left( M\lambda^a\lambda^a+M_\chi\chi^a\chi^a+ 2m_D\chi^a\lambda^a+h.c.\right)\nonumber
\\
&&-\sqrt{2}g_s\Big[ \tilde{d}_{Lxi}^*T^a_{xy}\lambda^{a\alpha}  d_{Lyi\alpha} -\tilde{d}_{Rxi}T_{xy}^{a*}\lambda^{a\alpha}d_{Ryi\alpha}^c\Big]+h.c.\,,
\eea
where $\lambda_\alpha^a$ is the Majorana gaugino, $\chi^a_\alpha$ its Dirac partner and $T^a_{xy}$, $d_{Li}$, $\tilde{d}_{Li}$ are the SU(3) generators, the quarks and the squarks of generation $i$ respectively\footnote{Our conventions are the ones from
\cite{drees}.}. The mass matrix is diagonalised by performing an orthogonal transformation and then a phase shift to render the masses positive,
\be
\left(\begin{array}{c}\lambda^a \\\chi^a\end{array}\right)= R \left(\begin{array}{c}\psi_1^a \\\psi_2^a\end{array}\right) .
\ee
In basis $\psi_i$, eq. (\ref{Ldirac}) becomes
\begin{align}\label{psi}
\cL'&\supset-{1\over 2}\left( M_1\psi_1^a\psi_1^a+M_2\psi_2^a\psi_2^a+h.c.\right)
\\
&-\sqrt{2}g_s\Big[ \tilde{d}_{Lxi}^*T^a_{xy}( R_{11}\psi_1^{a\alpha}\!+\!R_{12}\psi_2^{a\alpha} )  d_{Lyi\alpha} -\tilde{d}_{Rxi}T_{xy}^{a*}( R_{11}\psi_1^{a\alpha}\!+\!R_{12}\psi_2^{a\alpha} )d_{Ryi\alpha}^c\Big]\!\!+\!h.c.\nonumber
\end{align}
The four-fermion effective action is given by \cite{hagelin,masiero}
\be\label{effgen}
{\cal H}_{K}= \sum_{i=1}^5 C_i Q_i+\sum_{i=1}^3 \tilde{C}_i \tilde{Q}_i\,,
\ee
where the conventionally chosen basis of the dimension six operators is (now in Dirac notation)
\bea
&&Q_1=\ov{d}_x\gamma^\mu P_L s_x\,\ov{d}_n\gamma_\mu P_L s_n\,,\nonumber
\\
&&Q_2= \ov{d}_xP_L s_x\,\ov{d}_nP_L s_n\,,\nonumber
\\
&&Q_3= \ov{d}_xP_L s_n\,\ov{d}_nP_L s_x\,,\nonumber
\\
&&Q_4=\ov{d}_xP_L s_x\,\ov{d}_nP_R s_n\,,\nonumber
\\
&&Q_5= \ov{d}_xP_L s_n\,\ov{d}_nP_R s_x\,,
\eea
$\tilde{Q}_{1,2,3}$ are the R-projection analogues of $Q_{1,2,3}$ and
\bea\label{gencoefs}
C_1&=&ig_s^4W_{1K}W_{1L}\left( {11\over 36}|R_{1r}|^2|R_{1q}|^2\tilde{I}_4+{1\over 9}M_rM_qR_{1r}^{*2}R_{1q}^2I_4 \right) W_{K2}^\dag W_{L2}^\dag\,,\nonumber
\\
C_2&=&ig_s^4{17\over 18}W_{4K}W_{4L}I_4W_{K2}^\dag W_{L2}^\dag\,M_rM_qR_{1r}^{2}R_{1q}^2\,,  \nonumber
\\
C_3&=&-ig_s^4{1\over 6}W_{4K}W_{4L}I_4W_{K2}^\dag W_{L2}^\dag\,M_rM_qR_{1r}^{2}R_{1q}^2\,,  \nonumber
\\
C_4&=&ig_s^4W_{1K}W_{4L}\left({7\over 3}M_rM_qR_{1r}^{*2}R_{1q}^2I_4-{1\over 3}|R_{1r}|^2|R_{1q}|^2\tilde{I}_4\right)W_{K2}^\dag W_{L5}^\dag\nonumber
\\
&&-ig_s^4{11\over 18}W_{1K}W_{4L}\tilde{I}_4W_{K5}^\dag W_{L2}^\dag\,|R_{1r}|^2|R_{1q}|^2\,, \nonumber
\\
C_5&=&ig_s^4W_{1K}W_{4L}\left({1\over 9}M_rM_qR_{1r}^{*2}R_{1q}^2I_4+{5\over 9}|R_{1r}|^2|R_{1q}|^2\tilde{I}_4\right)W_{K2}^\dag W_{L5}^\dag\nonumber
\\
&&-ig_s^4{5\over 6}W_{1K}W_{4L}\tilde{I}_4W_{K5}^\dag W_{L2}^\dag\,|R_{1r}|^2|R_{1q}|^2\,,
\eea
\bea\label{gencoefs2}
&&\tilde{C}_1=ig_s^4W_{4K}W_{4L}\left( {11\over 36}|R_{1r}|^2|R_{1q}|^2\tilde{I}_4+{1\over 9}M_rM_qR_{1r}^{2}R_{1q}^{*2}I_4 \right) W_{K5}^\dag W_{L5}^\dag\,,\nonumber
\\
&&\tilde{C}_2=ig_s^4{17\over 18}W_{1K}W_{1L}I_4W_{K5}^\dag W_{L5}^\dag\,M_rM_qR_{1r}^{*2}R_{1q}^{*2}\,,  \nonumber
\\
&&\tilde{C}_3=-ig_s^4{1\over 6}W_{1K}W_{1L}I_4W_{K5}^\dag W_{L5}^\dag\,M_rM_qR_{1r}^{*2}R_{1q}^{*2}\,.
\eea
where the Feynman integrals\footnote{ See appendix \ref{defI} for explicit definitions.} are $I_4=I_4(M_r^2,M_q^2,m_K^2,m_L^2)$, $\tilde{I}_4=\tilde{I}_4(M_r^2,M_q^2,m_K^2,m_L^2)$ and summation over $r,q=1,2$ and $K,L=1,. . .\,,6$ is implied. $W_{IJ}$ is the unitary matrix that diagonalises the down squark mass-squared matrix $m^2_{\tilde{d}}$ in a basis where the down quark mass matrix is diagonal. Matrix $W$ is given in terms of the squark diagonalising matrix $Z$ and the quark diagonalising matrices $V_L$, $V_R$ by
\be\label{wvz}
W=\left(\begin{array}{cc} V_L^\dag Z_{LL} & V_L^\dag Z_{LR} \\V_R^\dag Z_{RL} & V_R^\dag Z_{RR}\end{array}\right)
\ee
as detailed in appendix \ref{feynrules}.

In the simple case that the mass of the gaugino is Dirac-type ($M=M_\chi=0$), we obtain $M_1=M_2=m_D$, $R_{11}\!=\!-iR_{12}\!=\!{1\over \sqrt{2}}$, so that $\sum |R_{1r}|^2|R_{1q}|^2\tilde{I}_4=\tilde{I}_4$ and $\sum M_rM_qR_{1r}^{*2}R_{1q}^2I_4\!=\!\sum M_rM_qR_{1r}^{2}R_{1q}^2I_4\!=\!\sum M_rM_qR_{1r}^{*2}R_{1q}^{*2}I_4=0$. The effective coefficients simplify to
\bea\label{diraccoefs}
&&C_1=ig_s^4{11\over 36}W_{1K}W_{1L}\tilde{I}_4 W_{K2}^\dag W_{L2}^\dag\,,\quad C_2=0\,,\quad C_3=0\,,  \nonumber
\\
&&C_4=-ig_s^4{1\over 3}W_{1K}W_{4L}\tilde{I}_4W_{K2}^\dag W_{L5}^\dag-ig_s^4{11\over 18}W_{1K}W_{4L}\tilde{I}_4W_{K5}^\dag W_{L2}^\dag\,\,, \nonumber
\\
&&C_5=ig_s^4{5\over 9}W_{1K}W_{4L}\tilde{I}_4W_{K2}^\dag W_{L5}^\dag-ig_s^4{5\over 6}W_{1K}W_{4L}\tilde{I}_4W_{K5}^\dag W_{L2}^\dag\,, \nonumber
\\
&&\tilde{C}_1=ig_s^4{11\over 36}W_{4K}W_{4L} \tilde{I}_4 W_{K5}^\dag W_{L5}^\dag\,,\quad\tilde{C}_2=0\,,\quad\tilde{C}_3=0\,,
\eea

The derivation of the effective action for the mixing between the other neutral mesons is the same as above. Therefore, the corresponding effective actions are given by simple substitution:
\bea
\cH_{B_d}&=&\cH_K(s\ra b,2\ra 3,5\ra 6)\,,\nonumber
\\
\cH_{B_s}&=&\cH_K(d\ra s,s\ra b,1\ra 2,2\ra 3,4\ra 5,5\ra 6)\,,\nonumber
\\
\cH_{D^0}&=&\cH_K(d\ra u,s\ra c,W\ra W^u)\,.
\label{EQ:KBDtranslation}\eea

\subsection{Flavour-violation observables}

Flavour violation in the Kaon mixing system is typically parametrised by the real and imaginary part of the mixing amplitude. These two are related to the mass difference between $K_L$ and $K_S$ and the CP violating parameter as
\be\label{deltaMK}
\Delta m_K=2\textrm{Re}\bra K^0|\cH_K|\ov{K}^0\ket
\,,\quad|\epsilon_K| =\left|\frac{\mathrm{Im} \bra K^0 |\cH_K| \ov{K}^0 \ket}{\sqrt{2} \Delta m_K} \right|\ , 
\ee
which have both been experimentally measured with great accuracy \cite{pdg}. Their size sets strict bounds on the amount of flavour violation allowed by new physics. In order to compute these observables we need to extract the hadronic matrix elements of the operators in (\ref{effgen}). They are first derived in the Vacuum Saturation Approximation (VSA),
\bea\label{vsa}
\bra K^0| Q_1 |\ov{K}^0\ket_{VSA}&=&{1\over 3} m_K f_K^2\,,\nonumber
\\
\bra K^0| Q_2 |\ov{K}^0\ket_{VSA}&=&-{5\over 24}\left( {m_K\over m_s+m_d} \right)^2 m_K f_K^2\,,\nonumber
\\
\bra K^0| Q_3 |\ov{K}^0\ket_{VSA}&=&{1\over 24}\left( {m_K\over m_s+m_d} \right)^2 m_K f_K^2\,,\nonumber
\\
\bra K^0| Q_4 |\ov{K}^0\ket_{VSA}&=&\left[{1\over 24}+{1\over 4}\left( {m_K\over m_s+m_d} \right)^2\right] m_K f_K^2\,,\nonumber
\\
\bra K^0| Q_5 |\ov{K}^0\ket_{VSA}&=&\left[{1\over 8}+{1\over 12}\left( {m_K\over m_s+m_d} \right)^2\right] m_K f_K^2\,.
\eea
Since only strong interactions are involved, we get identical expressions for the `R-projection' version of the first three operators. The ratio of the exact over the VSA result for each of the five operators above is parametrised by the ``bag'' factors $B_i$, $i=1, . . . , 5$ (see app. \ref{APP:INPUT}), that are typically extracted by numerical techniques. In comparing with the SM contribution, the usual parametrisation used is
\be
{\textrm{Re}\bra K^0|\cH_K|\ov{K}^0\ket\over \textrm{Re}\bra K^0|\cH_K^{SM}|\ov{K}^0\ket}=C_{\Delta m_K}\,,\quad {\textrm{Im}\bra K^0|\cH_K|\ov{K}^0\ket\over \textrm{Im}\bra K^0|\cH_K^{SM}|\ov{K}^0\ket}=C_{\epsilon_K} \ . 
\ee

Flavour violation in $B_q$ meson systems is parametrised in a similar way, by the modulus and the phase of the mixing amplitude:
\be
{\bra B_q^0|\cH_{B_q}|\ov{B}_q^0\ket\over \bra B_q^0|\cH_{B_q}^{SM}|\ov{B}_q^0\ket} \ = \ C_{B_q} \ e^{2i\phi_{B_q}} \ ,
\ee
where the $B_q$-meson hadronic matrix elements are obtained by eq. (\ref{vsa}) by substitution $(m_K,f_K,m_s,m_d)\rightarrow (m_{B_q},f_{B_q},m_b,m_q)$ and the corresponding bag factors (see appendix \ref{APP:INPUT}). Finally, there exists a similar parametrisation of the D-meson mixing CP conserving and violating parameters (as in e.g. \cite{Bergmann:2000id}) which we do not explicitly describe here, and will be mentioned
in the appropriate place in section \ref{SEC:Models}. 

\subsection{Flavour patterns}
\label{patterns}

The stringent experimental bounds on flavour violation processes require that contributions from extensions of the Standard Model be highly suppressed. This is typically achieved by employing particular patterns for the flavour structure of the BSM theory. In the following we describe how flavour violation is parametrised in the patterns that will appear throughout the paper.

\subsubsection*{Degeneracy - mass insertion approximation}

One way to suppress flavour violation is to assume that the masses of the squarks are almost degenerate, $m_{I}^2=m_{\tilde{q}}^2+\delta m_I^2$, where $m_I^2$ are the squark mass eigenvalues and $\delta m_I^2$ are small enough deviations from an ``average'' squark mass-squared $m_{\tilde{q}}^2$, $I=1,...,6$. Expansion of the loop integrals in $\delta m_I^2 $ and use of the unitarity of the $W$ matrices delivers (for $I\neq J$, $L\neq N$)
\be
W_{IK}W_{LM}I_4(m^2_K,m^2_M)W_{KJ}^\dag W_{MN}^\dag =
m^2_{IJ} m^2_{LN}I_6(m^2_{\tilde{q}},m^2_{\tilde{q}},m^2_{\tilde{q}},m^2_{\tilde{q}})+...
\ee
where $m^2$ is the squark squared mass matrix in the basis where the quark mass matrix is diagonal. Flavour violation in this scheme is parametrised by the small ratio of the off-diagonal elements $m^2_{IJ}$ over the average squark mass $\delta_{ij}^{L(R )L(R )} \equiv  m_{\tilde{q}}^{-2}m^2_{i(i+3)\,j(j+3)}$.

\subsubsection*{Hierarchy}

A slightly different notation is used in the case of hierarchical squark masses where the squarks of first and second generation are much heavier than those of the third so that their contribution to the box diagrams is negligible. Further below we will consider such flavour patterns, in the simpler case of absent left-right mixing. In this case, one can parametrise flavour violation processes by $\hat{\delta}_{ij}^L \equiv  W^L_{i\,3}\,  W^{L\dag}_{3\,j}$, $\hat{\delta}_{ij}^R \equiv  W^R_{i\,3}\,  W^{R\dag}_{3\,j}$, where $W^L_{ij}$ and $W^R_{ij}$ are the block diagonal matrices of (\ref{wvz}). The reasoning behind this choice can be illustrated by the following example \cite{gnr}. Let us assume that $\tilde{b}_L$ is much lighter than the other squarks. Then
\be
W_{1K}W_{1L}I_4(m_K^2,m_L^2)W^\dag_{K2}W^\dag_{L2}\ \simeq\ (\hat{\delta}_{12}^L)^2\,I_4(m_{\tilde{b}_L}^2,m_{\tilde{b}_L}^2)\quad \mathrm{where} \ (\hat{\delta}_{12}^L)=W^L_{13}W^{L\dag}_{32}\,.
\ee
 
\subsubsection*{Alignment}

An alternative to degeneracy or hierarchy for the suppression of flavour violating processes is to consider that the squark  mass-squared matrix is simultaneously diagonalised with the quark mass matrix \cite{nirseiberg}. In this ``alignment'' flavour pattern, the suppression appears because $W_{ij}^L=V_{ik}^{L\,\dag}Z^{LL}_{kj}\sim \delta_{ij}$ and $W_{ij}^R=V_{ik}^{R\,\dag}Z^{RR}_{kj}\sim \delta_{ij}$. In this framework, we can take the squark masses to be of the same order $m_{\tilde{q}}$ but not degenerate. If we ignore left-right mixing, we obtain e.g. for the left sector
\be
W^L_{1i}W^L_{1j}I_4(m^2_i,m^2_j)W^{L\dag}_{i2}W^{L\dag}_{j2}\simeq(\tilde{\delta}_{12}^L)^2I_4(m^2_{\tilde{q}},m^2_{\tilde{q}})\,\quad \mathrm{where}\ \tilde{\delta}^L_{12}= \underset{k}{\mathrm{max}}(W^L_{1k}W^{L\dag}_{k2})
\ee
and similarly for $\tilde{\delta}^R_{12}$.

\section{Bounds in the mass insertion approximation}
\label{SEC:MassInsertion}

In the following we present the bounds for representative points in the gluino parameter space $(M,m_D,M_\chi)$. We focus on near degenerate squarks; hierarchical and alignment flavour patterns are discussed in section \ref{SEC:Models}. In this approximation, coefficients (\ref{gencoefs}) and (\ref{gencoefs2}) of the general effective action for the Kaon mixing system become
\bea
&&C_1=-{\alpha_s^2\over m^2_{\tilde{q}}}\left( {11\over 36}|R_{1r}|^2|R_{1q}|^2\tilde{f}_6+{1\over 9}\sqrt{x_rx_q}R_{1r}^{*2}R_{1q}^2f_6 \right)(\delta_{12}^{LL})^2\,,\nonumber
\\
&&C_2=-{17\over 18}{\alpha_s^2\over m^2_{\tilde{q}}}\,\sqrt{x_rx_q}R_{1r}^{2}R_{1q}^2f_6(\delta_{12}^{RL})^2\,,  \nonumber
\\
&&C_3={1\over 6}{\alpha_s^2\over m^2_{\tilde{q}}}\,\sqrt{x_rx_q}R_{1r}^{2}R_{1q}^2f_6(\delta_{12}^{RL})^2\,,  \nonumber
\\
&&C_4=-{\alpha_s^2\over m^2_{\tilde{q}}}\Big[\Big({7\over 3}\sqrt{x_rx_q}R_{1r}^{*2}R_{1q}^2f_6-{1\over 3}|R_{1r}|^2|R_{1q}|^2\tilde{f}_6\Big)\delta_{12}^{LL}\delta_{12}^{RR}\nonumber
\\
&&\qquad\qquad\quad-{11\over 18}\,|R_{1r}|^2|R_{1q}|^2\tilde{f}_6\,\delta_{12}^{LR}\delta_{12}^{RL}\Big]\,, \nonumber
\\
&&C_5=-{\alpha_s^2\over m^2_{\tilde{q}}}\Big[\Big({1\over 9}\sqrt{x_rx_q}R_{1r}^{*2}R_{1q}^2f_6+{5\over 9}|R_{1r}|^2|R_{1q}|^2\tilde{f}_6\Big)\delta_{12}^{LL}\delta_{12}^{RR}\nonumber
\\
&&\qquad\qquad\quad-{5\over 6}\,|R_{1r}|^2|R_{1q}|^2\tilde{f}_6\,\delta_{12}^{LR}\delta_{12}^{RL}\Big]\,, \nonumber
\eea
\bea
&&\tilde{C}_1=-{\alpha_s^2\over m^2_{\tilde{q}}}\left( {11\over 36}|R_{1r}|^2|R_{1q}|^2\tilde{f}_6+{1\over 9}\sqrt{x_rx_q}R_{1r}^{2}R_{1q}^{*2}f_6 \right)(\delta_{12}^{RR})^2\,,\nonumber
\\
&&\tilde{C}_2=-{17\over 18}{\alpha_s^2\over m^2_{\tilde{q}}}\,\sqrt{x_rx_q}R_{1r}^{*2}R_{1q}^{*2}f_6(\delta_{12}^{LR})^2\,,  \nonumber
\\
&&\tilde{C}_3={1\over 6}{\alpha_s^2\over m^2_{\tilde{q}}}\,\sqrt{x_rx_q}R_{1r}^{*2}R_{1q}^{*2}f_6(\delta_{12}^{LR})^2\,,
\eea
while for the $B_d$ and $B_s$ system we replace $\delta_{12}\rightarrow \delta_{13}$ and $\delta_{12}\rightarrow \delta_{23}$ accordingly. In the expressions above, $x_k= M_k^2 / m_{\tilde{q}}^2$ with $M_k$ the gluino mass eigenstate and we have replaced,  according to appendix \ref{defI} notations with mass scale $m_{\tilde{q}}^2$, 
\bea
&&I_6(M_r^2,M_q^2,m_{\tilde{q}}^2,m_{\tilde{q}}^2,m_{\tilde{q}}^2,m_{\tilde{q}}^2)={i\over 16\pi^2 m_{\tilde{q}}^8}f_6(x_r,x_q,1,1,1)={if_6\over 16\pi^2 m_{\tilde{q}}^8}\nonumber\,,
\\
&&\tilde{I}_6(M_r^2,M_q^2,m_{\tilde{q}}^2,m_{\tilde{q}}^2,m_{\tilde{q}}^2,m_{\tilde{q}}^2)={i\over 16\pi^2 m_{\tilde{q}}^6}\tilde{f}_6(x_r,x_q,1,1,1)={i\tilde{f}_6\over 16\pi^2 m_{\tilde{q}}^6}\,.
\eea

The bounds on $d\leftrightarrow s$ transitions from the Kaon system are proven to be the most restrictive and therefore we will focus on them; we discuss the comparison of bounds in appendix \ref{APP:Bs}. We allow the SUSY contribution to $\Delta m_K$ to be as large as the experimental bound; however, the contribution to $\epsilon_K$ is restricted by the SM calculation \cite{utfit}. Our analysis takes into account NLO corrections to the effective Hamiltonian \cite{Ciuchini:1998ix}; as for the parameter inputs, they are given in appendix \ref{APP:INPUT}\footnote{Higher order terms in $B_4$ and $B_5$ of (\ref{vsa}) have been dropped \cite{Allton:1998sm}.}. 

\subsection{Majorana gluino}

In tables \ref{tblmajo2} and \ref{tblmajo1}, we update the bounds on flavour violation parameters for the MSSM with a Majorana gluino, for an average gluino mass of $1.5\,$TeV and $2\,$TeV. The results are identical for $\textrm{Re}(\delta^2)$ and $c^2\textrm{Im}(\delta^2)$, with\footnote{Saturating the $2\sigma$ deviation in $\epsilon_K^{SM}$.} $c \simeq 25$. As seen in the tables, the $K-\ov{K}$ system sets powerful constraints in the size of flavour violation. For example, for $m_{\tilde{q}}=2M_{\tilde{g}}=3\,$TeV the best case is $\sqrt{\textrm{Re}\,\delta^2}\lesssim 8\%$, while $\sqrt{\textrm{Im}\,\delta^2}$ is around 25 times smaller.

\begin{table}[htdp]
\begin{center}
\renewcommand{\arraystretch}{1.5}
\begin{tabular}{|c|c|c|c|}
\hline $m_{\tilde{q}}$ [GeV] & $\delta^{LL}\neq 0$ & $\delta^{LL}=\delta^{RR}\neq 0$ & $\delta^{LR}=\delta^{RL} \neq 0$ \\\hline 750 & 0.211 & 0.002 & 0.004 \\\hline 1500 & 0.180 & 0.002 & 0.014 \\\hline 2000 & 0.157 & 0.003 & 0.008 \\\hline  \end{tabular}
\end{center}
\caption{\small Majorana gluino bounds for $M_{\tilde{g}}=1500\,$GeV. By $\delta^{AB}$ we denote $\sqrt{|\textrm{Re}\,(\delta_{12}^{AB})^{2}|}$ and $c\sqrt{|\textrm{Im}\,(\delta_{12}^{AB})^{2}|}$.}
\label{tblmajo2}
\end{table}

\begin{table}[htdp]
\begin{center}
\renewcommand{\arraystretch}{1.5}
\begin{tabular}{|c|c|c|c|}
\hline $m_{\tilde{q}}$ [GeV] & $\delta^{LL}\neq 0$ & $\delta^{LL}=\delta^{RR}\neq 0$ & $\delta^{LR}=\delta^{RL} \neq 0$ \\\hline 750 & 0.192 & 0.002 & 0.005 \\\hline 1500 & 0.374 & 0.003 & 0.011 \\\hline 2000 & 0.240 & 0.003 & 0.019 \\\hline  \end{tabular}
\end{center}
\caption{\small Majorana gluino bounds for $M_{\tilde{g}}=2000\,$GeV. By $\delta^{AB}$ we denote $\sqrt{|\textrm{Re}\,(\delta_{12}^{AB})^{2}|}$ and $c\sqrt{|\textrm{Im}\,(\delta_{12}^{AB})^{2}|}$.}
\label{tblmajo1}
\end{table}

\subsection{Dirac gluino}

\begin{figure*}
\includegraphics[scale=0.85]{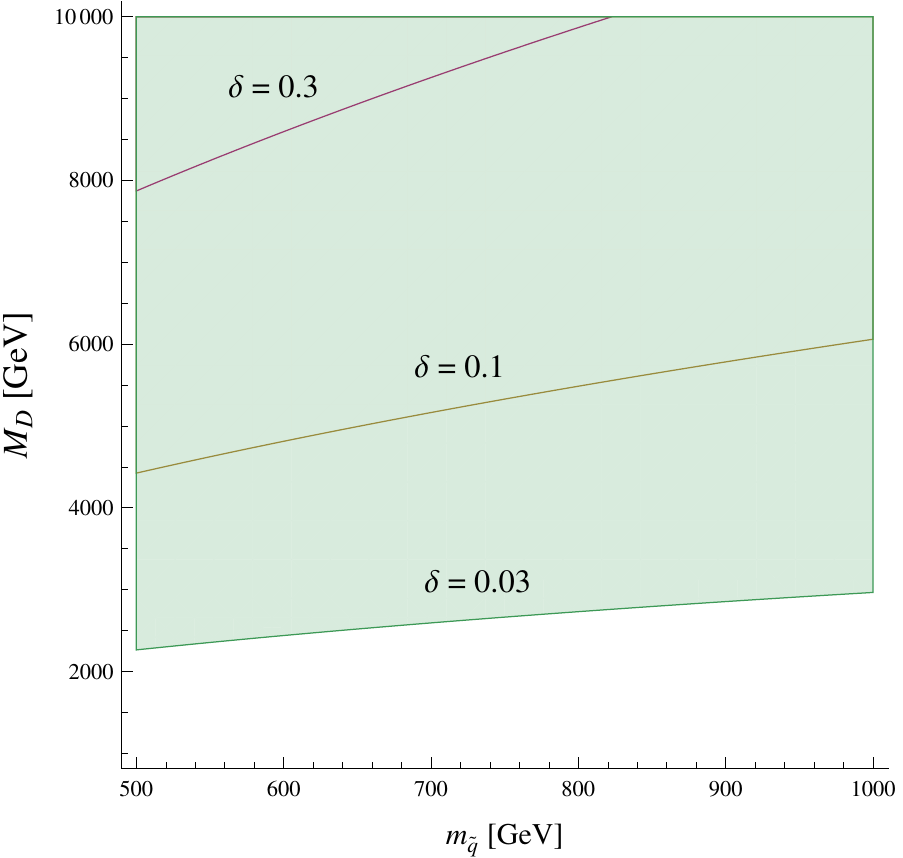}
\includegraphics[scale=0.85]{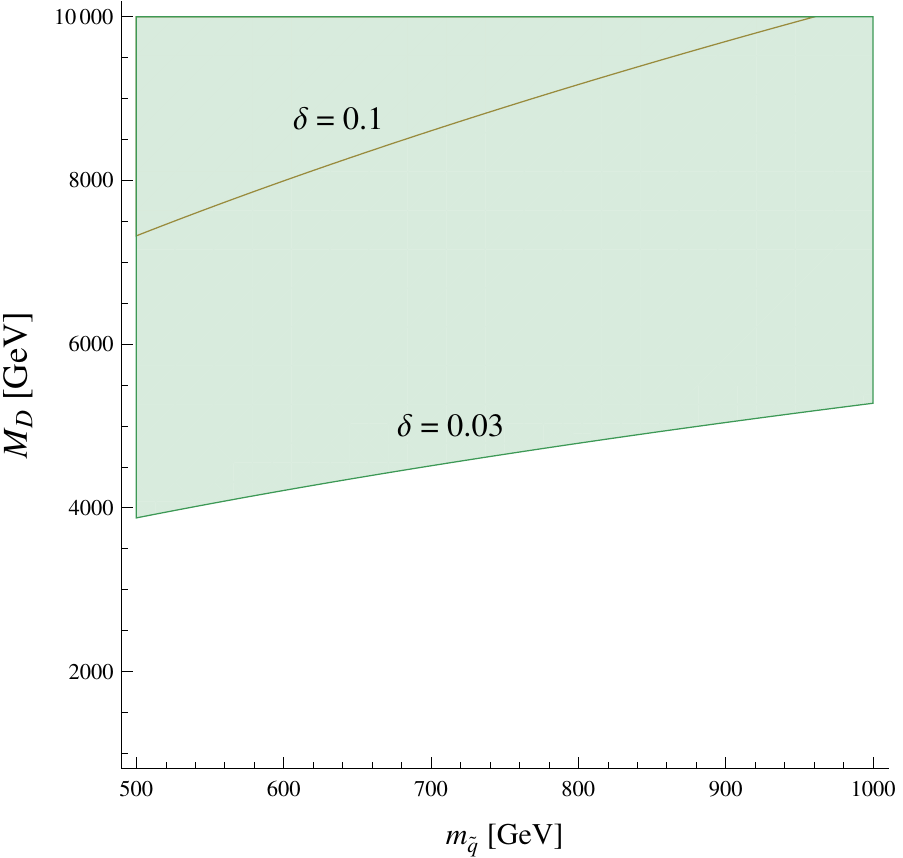}
\caption{\small Contour plots in parameter space $m_{\tilde{q}}$ - $m_D$ for purely Dirac gluino ($M=M_\chi=0$). Left: $\delta^{LL}=\delta^{RR}=\delta$, $\delta^{LR}=\delta^{RL}=0$. Right: $\delta^{LL}=\delta^{RR}=\delta^{LR}=\delta^{RL}=\delta$. Along the contours $\Delta m_K=\Delta m_K^{\textrm{exp}}$ (for $\delta^{AB}=\sqrt{|\textrm{Re}\,(\delta_{12}^{AB})^{2}|}$) and $\epsilon_K=\epsilon_K^{\textrm{exp}}$ (for $\delta^{AB}=c\sqrt{|\textrm{Im}\,(\delta_{12}^{AB})^{\,2}|}$).
}\label{diracLR}
\end{figure*}

As has already been mentioned in the introduction, flavour violation for quasi-degenerate squarks is suppressed if the gluino is of Dirac type, especially in the large gluino mass limit. This is true both because of the absence of the chirality-flip processes and because we are allowed to increase a Dirac gluino mass over the squark masses without affecting naturalness as much as in the Majorana case. These properties lead to a significant relaxation of the bounds from $\Delta m_K$ and $\epsilon_K$, as seen in figure \ref{diracLR} for representative values of $\delta^{AB}$.

However, despite the order of magnitude (or better) improvement over the Majorana case, the bounds on $\epsilon_K$ still require a relatively high flavour degeneracy or that the flavour violating masses in the squark matrix be real. For example, for a 6 TeV gluino and average squark mass of 1 TeV, $\sqrt{|\textrm{Im}\,(\delta_{12}^{LL})^2|}$ can be as high as $\sim 1\%$.

In section \ref{SEC:Models} we will explore flavour bounds on models with Dirac gauginos beyond the mass insertion approximation. We will see that there exist flavour models where a Dirac gluino can satisfy even the $\epsilon_K$ bounds for reasonable values of gluino and squark masses. We will also notice that in many other flavour models, Dirac gauginos do not enjoy the suppression of flavour violation with respect to Majorana ones that is seen here.

\subsection{Fake gluino}

The mass terms of eq. \ref{Ldirac} allow for non-standard gluinos, when all $M$, $m_D$ and $M_\chi$ are non-zero. One such scenario is when $M\gg M_\chi,m_D$ and corresponds to the interesting case of a light gluino with a suppressed squark - quark vertex, which we call ``fake gluino''. In section \ref{SEC:FakeGaugino} we explore this possibility in more detail.

In this limit we obtain much lower bounds on flavour violation parameters with respect to MSSM with Majorana gluino. In order to illustrate the point, we consider $m_D=M_\chi=M/10$. Even for an order of magnitude difference between $M$ and $m_D$, $M_\chi$, we obtain no restrictions for the size of flavour violation from effective operator $Q_1$, where $\delta^{LL}\neq 0$, $\delta^{RR}=\delta^{LR}=\delta^{RL}=0$. For other combinations, we obtain results given in tables \ref{tblwrong1} and \ref{tblwrong2}.

\begin{table}[htdp]
\begin{center}
\renewcommand{\arraystretch}{1.5}
\begin{tabular}{|c|c|c|}
\hline $m_{\tilde{q}}$ [GeV] & $\delta^{LL}=\delta^{RR}\neq 0$ & $\delta^{LR}=\delta^{RL} \neq 0$ \\\hline 750   & 0.013 & 0.028 \\\hline 1500  & 0.014 & 0.029 \\\hline 2000 & 0.014 & 0.030 \\\hline  \end{tabular}
\end{center}
\caption{``Fake'' gluino bounds for $M_{\tilde{g}}=1500\,$GeV. By $\delta^{AB}$ we denote $\sqrt{|\textrm{Re}\,(\delta_{12}^{AB})^{2}|}$ and $c\sqrt{|\textrm{Im}\,(\delta_{12}^{AB})^{2}|}$. }
\label{tblwrong1}
\end{table}
\begin{table}[htdp]
\begin{center}
\renewcommand{\arraystretch}{1.5}
\begin{tabular}{|c|c|c|}
\hline $m_{\tilde{q}}$ [GeV] & $\delta^{LL}=\delta^{RR}\neq 0$ & $\delta^{LR}=\delta^{RL} \neq 0$ \\\hline 750   & 0.017 & 0.037 \\\hline 1500  & 0.018 & 0.038 \\\hline 2000 & 0.018 & 0.039 \\\hline  \end{tabular}
\end{center}
\caption{``Fake'' gluino bounds for $M_{\tilde{g}}=2000\,$GeV. By $\delta^{AB}$ we denote $\sqrt{|\textrm{Re}\,(\delta_{12}^{AB})^{2}|}$ and $c\sqrt{|\textrm{Im}\,(\delta_{12}^{AB})^{2}|}$.}
\label{tblwrong2}
\end{table}

In this case, the quark/squark coupling of the fake gluino is suppressed with respect to the standard one by $R_{12} \sim \frac{m_D}{M} = 0.1$ as can be seen in eq. (\ref{psi}). So if the contribution to the box diagram is dominated by the lightest eigenstate, we should expect the box diagram to be suppressed by $R_{12}^4$ for the same lightest gluino mass, leading to bounds reduced by $R_{12}^2 \sim 0.01$. However, we observe from the bounds in tables \ref{tblwrong1} and \ref{tblwrong2} that the suppression is much less dramatic, of the order $0.1$. The reason is that it is not the light but actually the \emph{heavy} eigenstate that dominates the box integral!

This can be seen by comparing, for example, the loop integral contribution from the chirality-flip process:
\begin{align}
\sqrt{x_rx_q}R_{1r}^{*2}R_{1q}^2f_6(x_r,x_q) \simeq&\ x_1 f_6(x_1,x_1)+x_2 \left(\frac{x_2}{x_1} \right)^2 f_6(x_2,x_2)+2\sqrt{x_1x_2} \left(\frac{x_2}{x_1} \right) f_6(x_1,x_2) \nn\\
=&\ \frac{x}{y} f_6(x/y, x/y)+ xy^2  f_6(x,x)+ 2 x y \sqrt{y}   f_6(x/y, x)
\end{align}
where $x_1\simeq 100 x$, $x_2\simeq x$ with $x\equiv M_{\tilde{g}}^2/m_{\tilde{q}}^2$, $y \equiv \frac{x_2}{x_1} $ (for the  lightest gluino eigenstate) and we have replaced $R_{11}\simeq 1$, $R_{12}^2\simeq \frac{x_2}{x_1}$. Since $f_6(x/y,x) \sim  y^2 \log y, f_6(x/y,x/y) \sim \frac{y^2}{6x^2}$, the dominant contribution comes from the heavy gluino term $x_1f_6(x_1,x_1)$ and is given by
\be
\sqrt{x_rx_q}R_{1r}^{*2}R_{1q}^2f_6(x_r,x_q) \simeq \frac{1}{6x} \frac{M^2_D}{M^2}\,.
\ee
The parametric scaling of the \emph{bound} on $\delta^{AB}$ is then
\begin{align}
\frac{|\delta_{\mathrm{Majorana}}^{AB}|}{|\delta_{\mathrm{fake\ gluino}}^{AB}|} \sim \frac{M}{m_D} 
\end{align}
which is much less than the naive scaling of $\frac{M^2}{m_D^2}$. 

\section{Beyond the mass insertion approximation}
\label{SEC:Models}

Having established in the previous section that the bounds from $\epsilon_K$ do not allow flavour-generic models at LHC-accessible energies even in the case of Dirac gaugino masses, we are led to the conclusion that it is likely that we either require an accidental suppression of the mixing between the first two generations or we must impose some additional structure on the squark mass matrices. It is therefore important to consider flavour models. However, in doing so we invariably find that the mass insertion approximation is no longer valid: in fact, it is hard to find \emph{any} models in which it would actually apply. Hence, in this section we shall investigate the consequences - and the general bounds - when we go beyond the mass insertion approximation in the context of Dirac gauginos. 

One of the most important things that we find in the general case is that the much-vaunted suppression of $\Delta F=2$ FCNC processes is in general much less marked; in fact, for certain specific cases the Majorana case is actually less suppressed! We explain this in section \ref{SEC:argument}. In the remainder of the section we then discuss specific flavour models to illustrate the different types of behaviour. We shall consider:

\begin{itemize}
\item The simple case of non-degenerate but same order of magnitude squark masses, 
where \emph{alignment} applies. 
\item A simple flavour model realising such a spectrum. 
\item The general case of an inverted hierarchy between the first two squark generations and the third, \`a la reference \cite{gnr}. In addition to changing the gluino masses to Dirac type, we will update the bounds with the latest flavour data and also take into account the LHC bounds on squark and gaugino masses. 
\item Models where in addition to the first two generations of squarks, the third generation of \emph{right-handed} squarks is also heavy. These models provide a minimum of extra coloured particles available to the LHC. 
\item A flavour model realising the above, as given in \cite{Dudas:2013pja} but with Dirac gaugino masses. This model highly restricts the allowed flavour violation by imposing additional symmetries upon the first two generations. 
\end{itemize}

In the following, we ignore left - right squark mixing and define $W^L_{ij}=W_{ij}$ and $W^R_{ij}=W_{i+3\,j+3}$ for $i,j\leqslant 3$. We also define
\be\label{fab}
\tilde{f}^{AB}=W^A_{1i}W^B_{1j}\tilde{I}_4(m_D^2,m_{Ai}^2,m_{Bj}^2)W^{A\dag}_{i2}W^{B\dag}_{j2}\,,
\ee
where $A=L,R$. Then the effective action (\ref{effgen}) can be written as
\be\label{HKDirac}
\cH_K^{Dirac}=C_1Q_1+\tilde{C}_1\tilde{Q}_1+C_4Q_4+C_5Q_5
\ee
where the Dirac coefficients (\ref{diraccoefs}) are written as
\be
C_1={11\over 36}ig_s^4\tilde{f}^{LL}\,,\quad \tilde{C}_1={11\over 36}ig_s^4\tilde{f}^{RR}\,,\quad C_4=-{1\over 3}ig_s^4\tilde{f}^{LR}\,,\quad C_5={5\over 9}ig_s^4\tilde{f}^{LR}
\ee

\subsection{Dirac versus Majorana}
\label{SEC:argument}

In reference \cite{kpw}, it was argued that the absence of chirality-flip processes in the case of Dirac gluinos leads to a suppression in the contribution to the box diagram by a factor $x\equiv M_{\tilde{g}}^2 / m_{\tilde{q}}^2$ as the Dirac mass becomes larger than the squark masses. In the following we show that this is generally not true beyond mass insertion approximation and even when it is, the flavour bounds are often relaxed by a factor of few rather than being parametrically reduced.

This can be immediately seen by taking the large $x$ limit in the loop functions that appear in the coefficients (\ref{gencoefs}) of the general expression (\ref{effgen}) for $\Delta F=2$ FCNC processes. Taking for simplicity equal masses $m_{\tilde{q}}$ for the squarks in the loop, these functions are (see app. \ref{defI}): 
\begin{align}
M_{\tilde{g}}^2 I_4 (M_{\tilde{g}}^2,M_{\tilde{g}}^2, m_{\tilde{q}}^2,m_{\tilde{q}}^2) \equiv& \frac{i}{16 \pi^2 m_{\tilde{q}}^2} x f_4 (x) = \frac{i}{16 \pi^2 m_{\tilde{q}}^2} \bigg[ \frac{ 2 x(x-1)-x(x+1) \textrm{ln} (x)}{(1-x)^3} \bigg] \ , \nn\\
\tilde{I}_4 (M_{\tilde{g}}^2,M_{\tilde{g}}^2, m_{\tilde{q}}^2,m_{\tilde{q}}^2) \equiv& \frac{i }{16 \pi^2 m_{\tilde{q}}^2} \tilde{f}_4 (x) =\frac{i}{16 \pi^2 m_{\tilde{q}}^2} \bigg[ \frac{x^2-2 x \textrm{ln} (x)-1}{(1-x)^3}\bigg] \ . 
\end{align}
Function $\tilde{f}_4(x)$ appears in both Dirac and Majorana cases while $x f_4(x)$ appears only in the Majorana case, corresponding to the chirality-flip process. Notice that $x f_4 (x)$ is always positive, and $\tilde{f}_4 (x)$ always negative; moreover they have broadly similar values except near $x=0$; for example $f_4(1) = 1/6, \tilde{f}_4 (1) = -1/3$. As $x \rightarrow \infty$ the ratio between them tends to $-\textrm{ln}(x) + 2$, which is not the aforementioned enhancement by a factor of $x$.

\begin{figure}
\begin{center}
\includegraphics[width=0.4\textwidth]{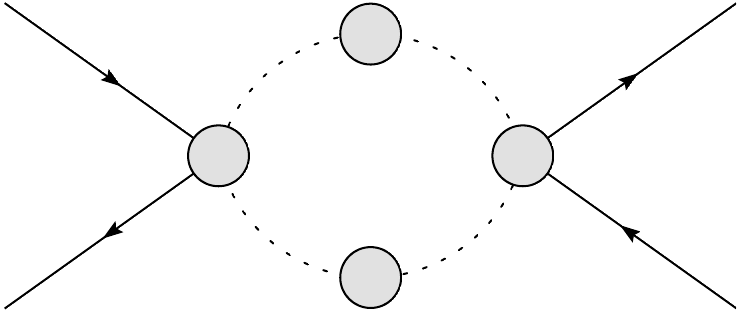} \includegraphics[width=0.4\textwidth]{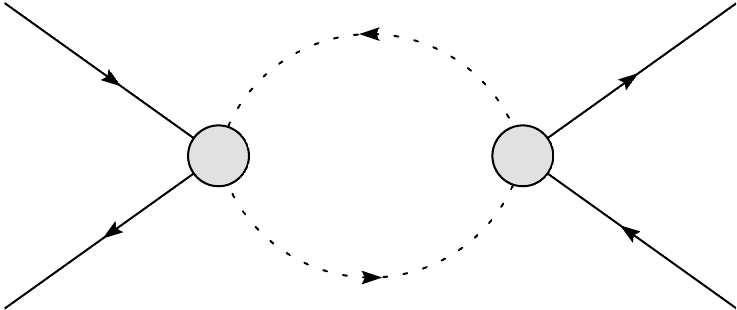}
\caption{Loop diagrams in the effective theory where the gaugino has been integrated out. In figure (a) the mass insertions are shown, whereas in figure (b) the mass-insertion approximation is inappropriate. } 
\label{FIG:Effectiveloops}
\end{center}
\end{figure}

This can be understood in the following way. Following the reasoning of \cite{kpw}, integrating out the heavy gluino generates effective operators
\be
\frac{1}{ M_{\tilde{g}}} \tilde{d}_R^* \tilde{s}_L^* \ov{d}_R s_L\,, \qquad  \frac{1}{M^2_{\tilde{g}}} \tilde{d}_L \partial_\mu \tilde{s}_L^* \ov{d}_L \gamma^\mu s_L\,,
\ee
the first of these being the chirality-flip process forbidden in the Dirac case. In the mass insertion approximation, the flavour changing loop diagram is then as in figure \ref{FIG:Effectiveloops}(a) and gives ($Q_i$ refers to the four-fermion effective operators of sec. \ref{SEC:Expressions})
\begin{align}
\mathcal{L}_{eff} \ \supset \ Q_i \  \frac{(m_{12}^2)^2}{ M_{\tilde{g}}^2}  \int d^4 q \frac{1}{(q^2 - m_{\tilde{q}}^2)^4} 
\ \sim \ Q_i \ \frac{\delta_{12}^2 }{ M_{\tilde{g}}^2}
\label{EQ:flipchiralmassinapprox}\end{align}
for the chirality-flip case and
\begin{align}
\mathcal{L}_{eff} \ \supset \ Q_i \  \frac{(m_{12}^2)^2}{ M_{\tilde{g}}^4 }  \int d^4 q \frac{q^2}{(q^2 - m_{\tilde{q}}^2)^4} 
\ \sim \ Q_i \ \frac{m_{\tilde{q}}^2}{ M_{\tilde{g}}^4}  \delta_{12}^2 
\label{EQ:samechiralmassinapprox}\end{align}
in the same chirality case, in line with the claim in  \cite{kpw}. The insertion of operators of the form $m_{12}^2 \tilde{q}_1^* \tilde{q}_2 $ as effective vertices is of course only valid in the limit $m_{12}^2 \ll m_{\tilde{q}}^2$; however, as we shall see below in section \ref{SEC:ALIGNMENT}, the above behaviour of the integrands can also arise in certain cases beyond mass insertion approximation, where there is approximate unitarity of a submatrix of the squark rotations leading to cancellations between diagrams. However, in all other cases we instead have diagrams like that of figure \ref{FIG:Effectiveloops}(b), which gives
\begin{align}
\mathcal{L}_{eff} \ \supset \ Q_i \  \frac{W_{12}^2}{ M_{\tilde{g}}^2}  \int^{M_{\tilde{g}}} \! d^4 q \frac{1}{(q^2 - m_{\tilde{q}}^2)^2}  \ \sim \ Q_i \ \frac{W_{12}^2}{ M_{\tilde{g}}^2}\, \textrm{ln} \frac{M_{\tilde{g}}^2}{m_{\tilde{q}}^2}
\label{EQ:flipchiralnonmassinapprox}\end{align}
in the chirality-flip case and
\begin{align}
\mathcal{L}_{eff} \ \supset \ Q_i \ \frac{W_{12}^2}{ M_{\tilde{g}}^4 } \int^{M_{\tilde{g}}} \! d^4 q \frac{q^2}{(q^2 - m_{\tilde{q}}^2)^2} 
\ \sim \  Q_i \ \frac{W_{12}^2}{ M_{\tilde{g}}^2}  
\label{EQ:samechiralnonmassinapprox}\end{align}
in the same chirality case, where we needed to use the cutoff of $M_{\tilde{g}}$ in the integrals\footnote{Note that if we define $m_{\tilde{q}_K}^2 = m_{\tilde{q}}^2 ( 1 + \delta_K)$, sum the integrals of the above form (\ref{EQ:flipchiralnonmassinapprox}) and (\ref{EQ:samechiralnonmassinapprox}) over $W_{1K} W_{2K}^* W_{1L} W_{2L}^*$ we and expand to leading order in $\delta_K$ we recover (\ref{EQ:flipchiralmassinapprox}) and (\ref{EQ:samechiralmassinapprox}).}. This is exactly the behaviour that we find born out in the amplitudes and explains why in generic flavour models the Dirac case will not provide a parametric suppression of the flavour-changing bounds.  

The logarithmic, instead of a linear suppression for the Dirac amplitude has then striking consequences. In the case that the contribution from same-chirality and chirality-flip amplitudes is comparable for reasonable values of $x$, the flavour bounds on Dirac gluinos can be proven more strict than those on Majorana, because in the latter there can exist cancellations between the same- and flipped- chirality amplitudes. Let us consider the impact that this has on bounds, by taking the ratio between the value of the Wilson coefficients $C_i$ for purely Majorana gauginos $C_i^M$ and for purely Dirac $C_i^D$. For a given contribution to the integrand (i.e. for the same values of $K, L$) in equation (\ref{gencoefs}) and taking for simplicity equal masses for the squarks in the loop (while neglecting left-right mixing) we find:
\begin{align}
{C_1^M \over C_1^D}=&1+{4\over 11}{xf_4(x,x,1)\over \tilde{f}_4(x,x,1)} =-{4\over 11}\ln(x) + {19\over 11}+\cO(x^{-1}\ln^2(x)), \nn\\
{C_4^M \over C_4^D}=& 7 \ln(x) -13 + \cO(x^{-1}\ln^2(x)), \nn\\
{C_5^M \over C_5^D}=& -\frac{1}{5} \ln(x) +\frac{7}{5} + \cO(x^{-1}\ln^2(x)).
\label{f4ratios}\end{align}
For arbitrarily large values of $x$ the Majorana case will have a larger contribution, but for reasonable values, up to $x=\mathcal{O}(100)$, only $C_4$ is actually enhanced compared to the Dirac case (for $C_1$ we would require gluinos about 40 times heavier than squarks to obtain a relative suppression). 

Finally, we note that the cancellation between the amplitudes can also be relevant when the the linear enhancement of the chirality-flip contribution applies, i.e. when $f^{AB}$ and $\tilde{f}^{AB}$ are proportional to $\overset{(\sim)}{I}_{\!\!\!6}$. This is the case when squarks are quasi-degenerate but also in certain cases beyond the mass insertion approximation for very particular squark matrix configurations as we shall find below. In this case, for moderate values of $x$ the cancellation plays a role:
\begin{align}
{C_1^M\over C_1^D}\rightarrow& 1+{4\over 11}{xf_6(x,x,1)\over \tilde{f}_6(x,x,1)}={1\over 11}(47-2x-12\ln(x))+\cO(x^{-1}\ln^2(x)), \nn\\
{C_4^M\over C_4^D}\rightarrow& \frac{7x }{2} -62 + 21 \ln x  +\cO(x^{-1}\ln^2(x)), \nn\\
{C_5^M\over C_5^D}\rightarrow& \frac{1}{10} (28 - x - 6 \ln x) +\cO(x^{-1}\ln^2(x))
\label{f6ratios}\end{align}
 We observe that the Majorana contribution is smaller than the Dirac for $C_1(x \lesssim 5)$ and $C_5(x \lesssim 15)$ while the Dirac is only suppressed by a factor of 10 for $C_1(x\simeq 50)$ and $C_5(x\simeq 100)$. 

\subsection{Alignment}
\label{SEC:ALIGNMENT}

In the previous section we examined how flavour constraints in the mass insertion approximation are affected by a generalised gluino spectrum. However, flavour models often do not lead to a near degeneracy of the squarks' masses but to different flavour patterns such as alignment or hierarchy, as mentioned in section \ref{patterns}. Moreover, one expects non-degeneracy to arise from running: there will always be a split between at least the first two generations and the third due to the larger Yukawa couplings. It therefore makes sense to consider models that can suppress flavour constraints even without requiring degeneracy of the squarks' masses.

\subsubsection*{Alignment in the left sector}

Alignment is typically obtained in flavour models of additional horizontal $U(1)$ symmetries \cite{Leurer:1993gy}. In a minimal representative of such models there is only one horizontal $U(1)$ symmetry, under which the quark superfields are charged with charges $X$ as
\begin{align}
X[Q_i] = (3,2,0)\,, \qquad X[\ov{U}_i] =(3,1,0)\,, \qquad X[\ov{D}_i] = (3,2,2)\,.
\end{align}
If we neglect D-term contributions to the squark masses, the order of magnitude structure of the squark mass matrices (before any quark rotations) is\footnote{In all flavour
abelian models in what follows, $\sim$ means order of magnitude only and not a precise
number. }
\begin{align}
m_{\tilde{d}_L}^2 \sim&\ m_F^2 \threemat[1,\epsilon,\epsilon^3][\epsilon,1,\epsilon^2][\epsilon^3,\epsilon^2,1], \qquad m_{\tilde{d}_R}^2 \sim m_F^2 \threemat[1,\epsilon,\epsilon][\epsilon,1,1][\epsilon,1,1].
\end{align}
where $\epsilon$ is a small number, the parameter of $U(1)$ symmetry breaking. Throughout this section, $\epsilon = \lambda$, where
$\lambda \simeq 0.22$ is the Cabibbo angle. In this flavour model, the quark diagonalising matrices have the same structure
\begin{align}
V_L^d \sim& \threemat[1,\epsilon,\epsilon^3][\epsilon,1,\epsilon^2][\epsilon^3,\epsilon^2,1], \qquad V_R^d \sim \threemat[1,\epsilon,\epsilon][\epsilon,1,1][\epsilon,1,1]
\end{align}
and the squark diagonalising matrices (in the basis where the quarks are diagonal) are approximated by $W^L\sim V_L^{d\,\dag}$ and $W^R\sim V_R^{d\,\dag}$. Therefore, with this particular choice of $U(1)$ charges, the left-squarks sector exhibits alignment while the right-squarks sector does not.

We can estimate flavour violation in $\Delta m_K$ in the leading order in $\epsilon$, by focusing at

\bea
\tilde{f}^{LR}&=&\epsilon^2\Big[(m_{R1}^2-m_{R2}^2)\left(\tilde{I}_5(m_{L1}^2,m_{R1}^2,m_{R2}^2)-\tilde{I}_5(m_{L2}^2,m_{R1}^2,m_{R2}^2)
\right) \ , \nonumber \\
&&\qquad+(m_{L2}^2-m_{L1}^2)\tilde{I}_5(m_{L1}^2,m_{L2}^2,m_{R3}^2)\Big]+\cO(\epsilon^4) \
\sim \ {i\over 16\pi^2}{\epsilon^2\over m_{\tilde{q}}^2}\tilde{f}_5(x) \ , \nn
\\
\eea
\bea
\tilde{f}^{LL}&=&\epsilon^2(m_{L1}^2-m_{L2}^2)^2\tilde{I}_6(m_D^2,m_{L1}^2,m_{L2}^2)+\cO(\epsilon^4) \
 \sim \ {i\over 16\pi^2}{\epsilon^2\over m_{\tilde{q}}^2}\tilde{f}_6(x) \ , \nn
\\
\tilde{f}^{RR}&=&\epsilon^2\Big[\sum_i\tilde{I}_4(m_{Ri}^2,m_{Ri}^2)-2\tilde{I}_4(m_{R1}^2,m_{R2}^2)-2\tilde{I}_4(m_{R1}^2,m_{R3}^2)+2\tilde{I}_4(m_{R2}^2,m_{R3}^2)\Big]\nonumber
\\
&& \ + \ \cO(\epsilon^4) \
\sim \  {i\over 16\pi^2}{\epsilon^2\over m_{\tilde{q}}^2}\tilde{f}_4(x)\, ,
\eea
where $x=m_D^2/m_{\tilde{q}}^2$ and in approximating, we have required that all squark masses are of the same order $m_{\tilde{q}}$ but not degenerate. In the limit of Dirac gluinos much heavier than $m_{\tilde{q}}$ we obtain 
\be
\frac{\bra K^0 | H_{eff} | \ov{K}^0 \ket}{\Delta m_K(\mathrm{exp})} \simeq  \left(\frac{\alpha_s}{0.1184}\right)^2 \left( \frac{15\ \mathrm{TeV}}{m_D} \right)^2 
e^{2 i \phi_K} \ , 
\ee
which is much too large: in order to meet the bounds from $\epsilon_K$ we would need $m_D \sim \C{O}(100)$ TeV. Here, we might have expected Dirac gaugino masses to soften the bounds with respect to Majorana masses. However, this is not the case. Since the strongest constraint comes from operator $\tilde{Q}_1$, according to equation (\ref{f4ratios}) we have a bound about $5$ times \emph{stronger} for Dirac masses than Majorana ones when $x=100$.

\subsubsection*{Alignment in both left and right sectors}

As we have seen above, since the constraints are severe for Kaon mixing, models that suppress the elements $W_{12}^L$ and $W_{12}^R$ are then most attractive (since $\tilde{f}^{AB}$ obtains largest contribution from $W^A_{11} W^{A\dag}_{12} \sim W^A_{12}$ and $W^A_{21} W^{A\dag}_{22} \sim W^A_{21}$). However, retrieving the correct form for the CKM matrix leads to large flavour rotation for the up-quark matrix. Therefore, apart from checking that B-meson constraints are satisfied, one must as well consider constraints from D-meson mixing.

Since both down and up squark sectors are involved in the following discussion, we restore the corresponding superscripts in the $W$ matrices, so that $W_{ij}^{q_A}$ is the matrix that diagonalises the $A$-handed squarks in the $q$-type sector, with $A=L,R$ and $q=u,d$.

Defining $\bra W_{ij}^q \ket \equiv \sqrt{W_{ij}^{q_L} W^{q_R}_{i,j}}$ we can place approximate bounds in this framework
\begin{align}
&W^{d_L}_{12}, W^{d_R}_{12} \lesssim 2\times 10^{-3}\,, \qquad \bra W_{12}^d \ket \lesssim 4 \times 10^{-4}\,, \nn
\\
&W^{d_L}_{13}, W^{d_R}_{13} \lesssim 0.1\,, \qquad\qquad\ \, \bra W_{13}^d \ket \lesssim 0.2\,, \nn
\\
&W^{d_L}_{23}, W^{d_R}_{23} \lesssim 0.4\,, \qquad\qquad\ \, \bra W_{23}^d \ket \lesssim 0.5\,, \nn
\\
&W^{u_L}_{21}, W^{u_R}_{21} \lesssim 0.03\,, \qquad\qquad\! \bra W_{21}^u \ket \lesssim 0.04\,,
\label{dmesoncons}
\end{align}
where all of these should be multiplied by $\left( \frac{m_{\tilde{q}}}{2\ \mathrm{TeV}} \right)  \sqrt{\left|\frac{1/3}{\tilde{f}_4 (x)}\right|} $.  The constraints\footnote{These approximate bounds include bag factors but no NLO corrections (no magic numbers) (in plots we include all available data including magic numbers).} in the left column of (\ref{dmesoncons}) come from operators of the type $Q_1, {\tilde Q}_1$, whereas the ones in the right column come from $Q_4, Q_5$.

Of these bounds, it is the D-meson constraint that proves problematic for alignment models, as typically suppressing the $W_{12}^d$ element will require $W^u_{21} \sim \lambda $. However, the problem is not particularly severe: it can either be remedied by having somewhat heavy first two generations, or by allowing a small degeneracy between the first two generations.

To explore this, consider as a representative example a model with two abelian symmetries $U(1)_1 \times U(1)_2$ under which the quark superfields have charges \cite{Leurer:1993gy} 
\begin{align}
\begin{array}{|c|c|c|}\hline
Q & \ov{D} & \ov{U} \\ \hline
(3,0) & (-1,2) & (-1,2) \\
(0,1) & (4,-1) & (1,0) \\
(0,0) & (0,1) & (0,0) \\\hline
\end{array}
\end{align}
Other examples of models with alignment can be found, e.g., in \cite{perez1,perez2}. The symmetry breaking parameters, coming from flavon fields of charges $(-1,0)$ and
$(0,-1)$, are $\epsilon_1 \sim \lambda$ and  $\epsilon_2 \sim \lambda^2$ respectively. The diagonalising matrices are given by
\begin{align}
W^{d_L}_{ij} \sim& \threemat[1,\lambda^5,\lambda^3][\lambda^5,1,\lambda^2][\lambda^3,\lambda^2,1], \qquad W^{d_R}_{ij} \sim \threemat[1,\lambda^7,\lambda^3][\lambda^7,1,\lambda^4][\lambda^3,\lambda^4,1] \ , \nn
\\
W^{u_L}_{ij} \sim& \threemat[1,\lambda,\lambda^3][\lambda,1,\lambda^2][\lambda^3,\lambda^2,1], \qquad W^{u_R}_{ij} \sim \threemat[1,\lambda^6,\lambda^5][\lambda^6,1,\lambda][\lambda^5,\lambda,1] \ , 
\end{align}
which are generically challenged by the bounds given above via $D$-meson mixing. However, those bounds are derived under the assumption that the amplitude is well dominated by a single contribution. We find that, in practice, they are overly conservative. Indeed, in order for this to be the case there has to actually be a substantial hierarchy between the squark masses, and then since there is a minimum mass for the second generation via LHC bounds we will find that the model will be less constrained than feared. Considering this model, the constraint essentially comes from the $Q_1$ operator for $D$-meson mixing. Moreover, if we were to suppress the amplitude by $\mathcal{O}(\lambda^2)$ then we would easily meet the constraints; hence we must only suppress the leading order contribution in $\lambda$, which we find to be:
\bea
\tilde{f}^{LL}& =& \lambda^2 \bigg[ \tilde{I}_4 (m_{L1}^2 ,m_{L1}^2) +  \tilde{I}_4 (m_{L2}^2 ,m_{L2}^2) - 2 \tilde{I}_4 (m_{L1}^2 ,m_{L2}^2) \bigg] +\cO(\lambda^4)\nn
\\
&=& \lambda^2 (m_{L1}^2 - m_{L2}^2)^2 \tilde{I}_6 (m_{L1}^2 ,m_{L2}^2 )+\cO(\lambda^4) .
\label{EQ:alignapprox}\eea
Clearly if the first two generations are quasi-degenerate then this will vanish sufficiently to satisfy the constraints. Indeed, particular UV models could have them degenerate up to $\mathcal{O}(\lambda^2)$ \cite{dgps}, which would give a much greater suppression of the FCNC processes than necessary to avoid current bounds. However, it is actually not necessary to have so much degeneracy; for example taking $m_{L1}^2 = 3 m_{L3}^2, m_{L2}^2 = 2 m_{L3}^2$ and taking $m_D = m_{L2} $ the amplitude is suppressed by a factor of $0.02$ compared to simply taking $\tilde{f}_4 (1)$, which is enough to satisfy the bounds for squark at gluino masses of $\mathcal{O}(2\ \mathrm{TeV})$.

\begin{figure}\begin{center}
\includegraphics[width=0.85\textwidth]{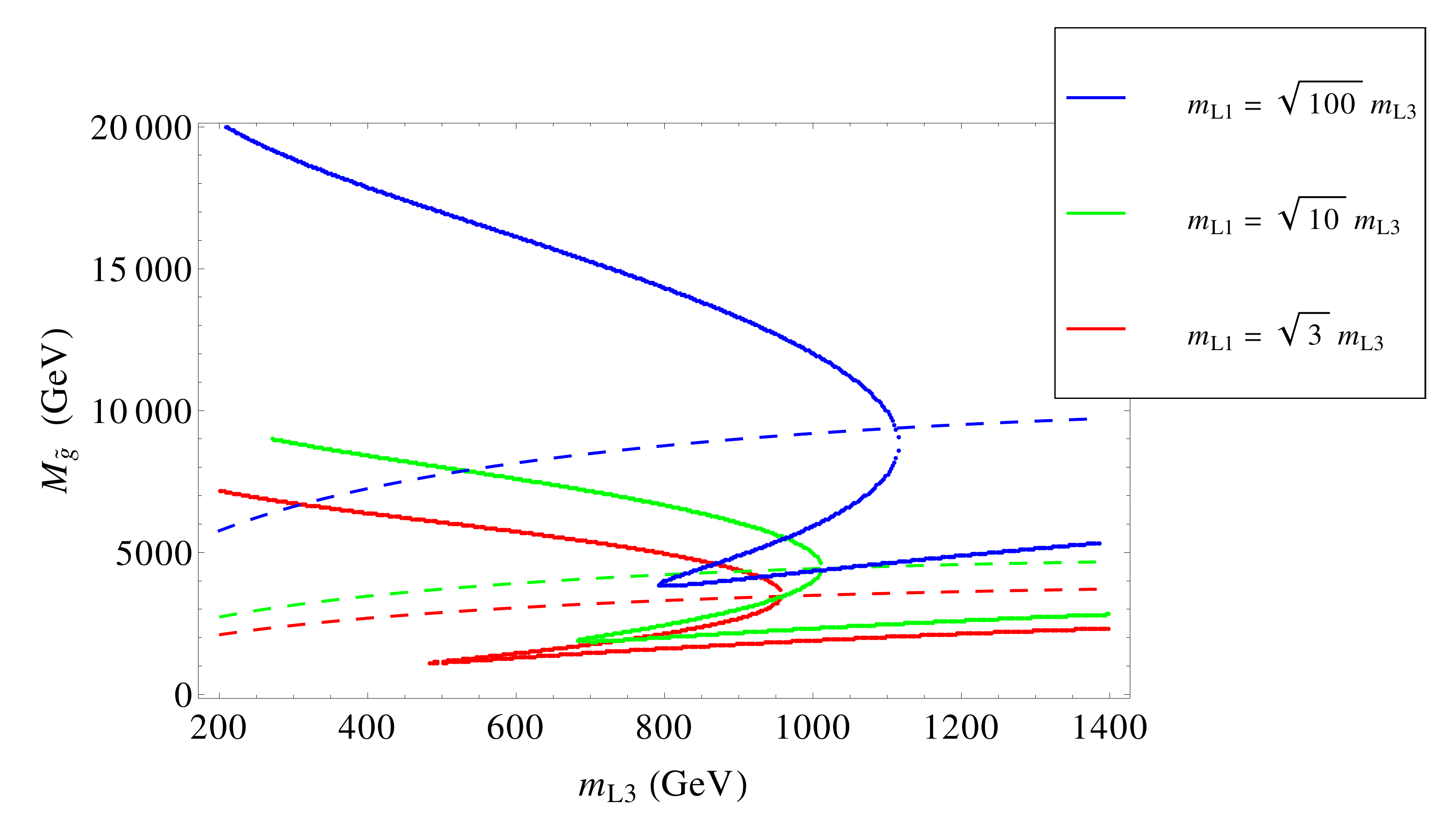}
\caption{Constraints on the model described in section \ref{SEC:ALIGNMENT}. The dashed lines correspond to exactly Dirac gauginos, while the solid lines are purely Majorana. We take the same hierarchies for up- and down-type squarks, with $m_{L1}^2 = 1.5 m_{L2}^2 = 3 m_{L3}^2$ for the red plots; $m_{L1}^2 = 5 m_{L2}^2 = 10 m_{L3}^2  $ for the green curves and $m_{L1}^2 = 25 m_{L2}^2 = 100 m_{L3}^2$ for the blue.}
\label{FIG:NirSeiberg}
\end{center}
\end{figure}

To illustrate this, we show plots in figure \ref{FIG:NirSeiberg} of the allowed lightest squark mass versus gaugino mass for this model with randomly chosen entries of the above form. In order to harden the bounds we must introduce a large hierarchy between the squark masses. We take three different hierarchies: $m_{L1}^2 = 1.5 m_{L2}^2 = 3 m_{L3}^2$,  $m_{L1}^2 = 5 m_{L2}^2 = 10 m_{L3}^2  $ and $m_{L1}^2 = 25 m_{L2}^2 = 100 m_{L3}^2$ (the same hierarchies for both up- and down-type squarks) and calculate the bounds showing the gluino mass against the \emph{lightest} squark mass  using NLO corrections and taking into account all $\Delta F =2$ constraints. In practice, the D-meson constraint is dominant: we insist that $|\Delta m_{D^0}|$ is less than the experimental value of $7.754 \times 10^{-15} $ GeV (since this is approximately three standard deviations from zero, and moreover the standard model value is known to much less accuracy). 

The results of figure \ref{FIG:NirSeiberg} agree with our discussion in the end of sec. \ref{SEC:argument}. The cancellation between the chirality-flip and the same chirality process suppresses the contribution in the Majorana case for moderate $x$ even if the enhancement over the Dirac case is linear in $x$. Since the flavour bounds for $m_{L3}\gtrsim 0.8 \div1\,$TeV are obeyed already at low $x$, a Majorana gluino is less constrained than a Dirac one. Another feature of this model is that, due to the suppression in the unitary rotations, the main FCNC effects come from the first two generations, even if they are heavier than the third one. Hence, one should bear in mind that the relevant squark mass for the loop diagrams is heavier than the $m_{L3}$ shown on the abscissa. 
\subsection{Inverted hierarchy}

\subsubsection{Decoupling the first two generations}

A particularly attractive scenario in light of the strong LHC bounds on the first two generations of squarks and the desire for ``natural supersymmetry'' is to have an inverted hierarchy, where the first two generations of squarks are substantially heavier than the third. This can be simply accommodated in flavour models, as we shall discuss below. 

One approach, following \cite{gnr}, is to decouple the first two generations. In this case, the effective action is given by (\ref{HKDirac}) with $\tilde{f}^{AB}$ of (\ref{fab}) given by
\be
\tilde{f}^{AB}=\hat{\delta}_{12}^A\,\hat{\delta}_{12}^B\,\tilde{I}_4(m_D^2,m_{A3}^2,m_{B3}^2)
\ee
in the inverted hierarchy limit, as we have described in sec. \ref{patterns}. Here $m_{L3}, m_{R3}$ are the masses of the `left-handed' and `right-handed' sbottoms. The reader should be careful with the ``hat'' notation however: since $\hat{\delta}^A_{12} \equiv W_{13}^A \ov{W}_{23}^A$ we expect the $\hat{\delta}_{12}^A$ to be small, coming from two small rotations rather than (in the generic case) one - indeed if the rotations come from the squark mass-squared matrices $M_{A\, ij}^2$ themselves (rather than from quark rotations) so that $W_{13}^A \simeq - M_{A\, 13}^2/m_{A1}^2 $  then we expect $\hat{\delta}_{12}^A < m_{A3}^2/m_{A1}^2$.

For $m_D^2 \gg m_{L3}^2,m_{R3}^2$  we find (we discuss the limits from $B$-meson mixing in appendix \ref{APP:Bs})
\be
\frac{\bra K^0 | H_{eff} | \ov{K}^0 \ket}{\Delta m_K (\mathrm{exp})} =3\times 10^3 \times \left(\frac{\alpha_s}{0.1184}\right)^2 \left( \frac{2000\, \mathrm{GeV}}{m_D}\right)^2 \left( 0.3(\hat{\delta}_{12}^{L})^2+0.3(\hat{\delta}_{12}^{R})^2-2.6\, \hat{\delta}_{12}^{L}\hat{\delta}_{12}^{R} \right) 
\ee
and hence
\be
\sqrt{|\mathrm{Re}(\hat{\delta}_{12}^{L})^2|} < 3 \times 10^{-2} \left( \frac{m_D}{2000\ \mathrm{GeV}} \right)\,,\quad \sqrt{|\mathrm{Im}(\hat{\delta}_{12}^{L})^2|} < 9 \times 10^{-4} \left( \frac{m_D}{2000\ \mathrm{GeV}} \right)\,,
\label{EQ:decoupledlimits}\ee
which are not much weaker than the limits from \cite{gnr} despite the larger gaugino mass and the change from Majorana to Dirac gauginos. The reason is that the flavour data have been updated and the limits scale only inversely proportional to the gaugino mass, since there is no further suppression of the Dirac case relative to the Majorana case, as described in section \ref{SEC:argument}. In fact, since the limits are derived from the constraints on $C_1, \tilde{C}_1$ without the mass insertion approximation, for moderate values of the ratio of gluino to third generation squark masses, the Dirac version of this model is actually more constrained than the Majorana one.

\subsubsection{Including the first two generations}


The above discussion assumed that we could completely decouple the first two generations. However, we know that we cannot make them arbitrarily massive compared to the third generation without the two-loop RGEs leading either to tachyons or substantial fine-tuning to avoid them. Typically a factor of $m_1/m_3 \sim 10-15$ is the maximum that is allowed. Given this, we must still worry about flavour-changing effects from the first two generations. 

For example, let us suppose that the heavy eigenstates are not degenerate, but  have masses $m_1 \leqslant m_2$. In the limit where $m_1$ is much larger than $m_D$, one of the contributions to $\tilde{f}^{LR}$ of (\ref{fab}) can be written as  
\be
-16\pi^2 i \tilde{f}^{LR} \sim \frac{W^L_{12}W^R_{12}}{m_1^2}\,.\nonumber
\ee
Under the reasonable assumption that there are no accidental cancellations between the different contributions, for $m_1 \sim 10\,$TeV the constraint from $\epsilon_K$  requires $W^L_{12}W^R_{12} \lesssim 10^{-6}$  which is clearly highly restrictive for any flavour model. Therefore we must impose restrictions upon the heavy squarks.

Let us determine the condition for neglecting the contribution from the first two generations in the approximation that the first two generations of left- and right-handed squarks are degenerate to leading order with masses $m_{L1}, m_{R1}$ respectively, with the third generation masses $m_{L3}, m_{R3}$. Then, there are corrections $\delta_{12}^{A} m_1^2$, $\delta_{13}^{A}  m_1^2$, $\delta_{23}^{A}  m_1^2$ to the off-diagonal elements of the squark mass-squared matrix, with $\delta_{ij}^A$ defined similar to the mass insertion approximation flavour parameter described in sec. \ref{patterns}: $\delta_{ij}^A= m_1^{-2} (m^{A}_{ij})^2$, $A=L,R$. In this case, eq. (\ref{fab}) is expressed as
\bea\label{EQ:hierarchy}
\tilde{f}^{AB} &\simeq& \hat{\delta}_{12}^A\,\hat{\delta}_{12}^B\, \tilde{I}_4 (m_{A3}^2, m_{B3}^2) \nn
\\
&&+\,\delta_{12}^A\,\hat{\delta}_{12}^B\,m_{A1}^2 \,\frac{\partial}{\partial m_{A1}^2} \bigg[  \tilde{I}_4 (m_{A1}^2, m_{B3}^2) - \tilde{I}_4 (m_{A1}^2, m_{B1}^2) \bigg] + (A \leftrightarrow B)\nn
\\
&&+\,\delta_{12}^A \delta_{12}^B\,m_{A1}^2\,m_{B1}^2\frac{\partial^2\tilde{I}_4 (m_{A1}^2, m_{B1}^2)}{\partial m_{A1}^2\partial m_{B1}^2} \,,
\eea
where we have neglected subleading terms in $\hat{\delta}^{A,B}_{12}$. If we further take $m_{A1} = m_{B1} =m_1$, $m_{A3} = m_{B3} =m_3 $, then this simplifies to 
\bea
-16\pi^2 i \tilde{f}^{AB}&=&\hat{\delta}_{12}^A \hat{\delta}_{12}^B \frac{1}{m_3^2}\tilde{f}_4 (\frac{m_D^2}{m_3^2})+ \,\delta_{12}^A \delta_{12}^B\frac{1}{m_1^2} \tilde{f}_6 (\frac{m_D^2}{m_1^2})\nn
\\
&-&\left[\delta_{12}^A\,\hat{\delta}_{12}^B \frac{1}{m_1^2} \tilde{f}_5({m_D^2\over m_1^2},{m_3^2\over m_1^2})+ (A \leftrightarrow B)\right]\,,
\eea
where
\be\label{ftilde5}
\tilde{f}_5({m_D^2\over m_1^2},{m_3^2\over m_1^2})=\log\frac{m_D^2}{m_1^2} + \frac{2m_D^4 - 3 m_D^2 m_3^2 + m_3^4 (1 + \log \frac{m_3^2}{m_D^2})}{(m_3^2 - m_D^2)^2}+\cO(m_1^{-2})\,.
\ee
Assuming that  $m_D \ll m_1$, in order to neglect the contribution of the first two generations we require $\delta_{12} \lesssim \hat{\delta}_{12}\frac{m_1}{m_3}$. Since, as explained above, we expect $\frac{m_1}{m_3} \lesssim 10 \div 15$,  we see that only certain flavour models will actually allow this.    

\subsubsection{Concrete Realisations}
\label{SEC:DudasModel}

In order to realise a model with heavy first two generations of squarks with suppressed mixing between them, we could consider models with a large D-term for an extra abelian  gauged flavour symmetry under which only the first two generations are charged, and obtain a natural supersymmetric spectrum \cite{ckn}. These D-term contributions were argued to be naturally generated (at least) in effective string models \cite{bddp}, to be positive and, in certain circumstances, to be dominant over the F-term contributions. It is then clear from (\ref{abelian3}) and (\ref{abelian4}) that precisely because the first generations of \emph{fermions} are lighter than the third one, the corresponding scalars are {\it predicted to be heavier}.  While such models would be one approach to realising the scenario of the previous subsection, there is currently no extant example that solves the FCNC problem of mixing between the first two generations (owing to the need to have degeneracy between them).

Another class of flavour models adds extra symmetry between the first two generations \cite{u2models,flavouredgauge}. In this case, we can effectively take the squark mass matrix to be diagonal, with flavour-changing processes only induced by the quark rotations combined with the (possibly small) non-degeneracies in the squark matrix (of course, if the squarks were degenerate then the super-GIM mechanism would lead to vanishing of the flavour-changing effects).

Taking the model of \cite{Dudas:2013pja} for $m_{L1}^2=m_{L2}^2 = m_1^2 \gg m_{L3}^2$, $m_{R1}^2=m_{R2}^2 = m_1^2 \simeq m_{R3}^2$ as an illustrative example of this scenario (see appendix \ref{non-ab} for more details), we have approximately
\be
\tilde{f}^{LR} \simeq {i\over 16\pi^2 }W^L_{13}W^{L\dag}_{32}W^R_{13}W^{R\dag}_{32}\frac{m_1^2 - m_{3R}^2}{m_1^4} \tilde{f}_5 ({m_D^2\over m_1^2}, {m_{L3}^2\over m_1^2}) \ ,
\ee
where $\tilde{f}_5$ is given in (\ref{ftilde5}) and the diagonalising matrices are given in terms of parameters of the model:
\be
W^L_{13}W^{L\dag}_{32}W^R_{13}W^{R\dag}_{32}=-s_d^2{m_d\over m_s}|V_{23}^d|^2e^{-2i\tilde{\alpha}_{12}}\,,
\ee
with $s_d$ and $V_{23}^d$ that take values $s_d^2 \simeq 0.2$ and $V_{23}^d \simeq 0.04$ in the best fit of one of the models in \cite{Dudas:2013pja}.

\begin{figure}\begin{center}
\includegraphics[width=0.5\textwidth]{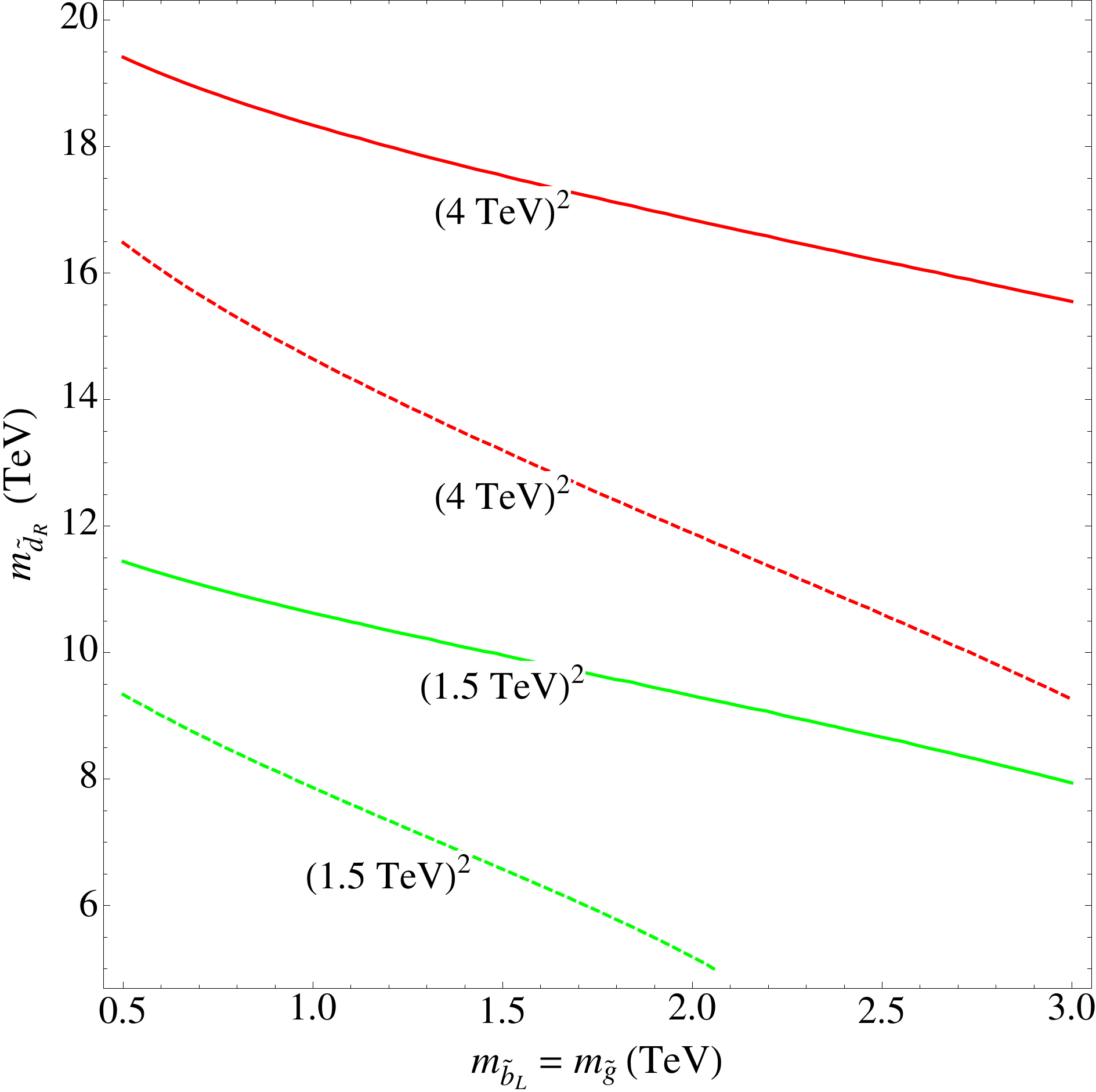}\includegraphics[width=0.5\textwidth]{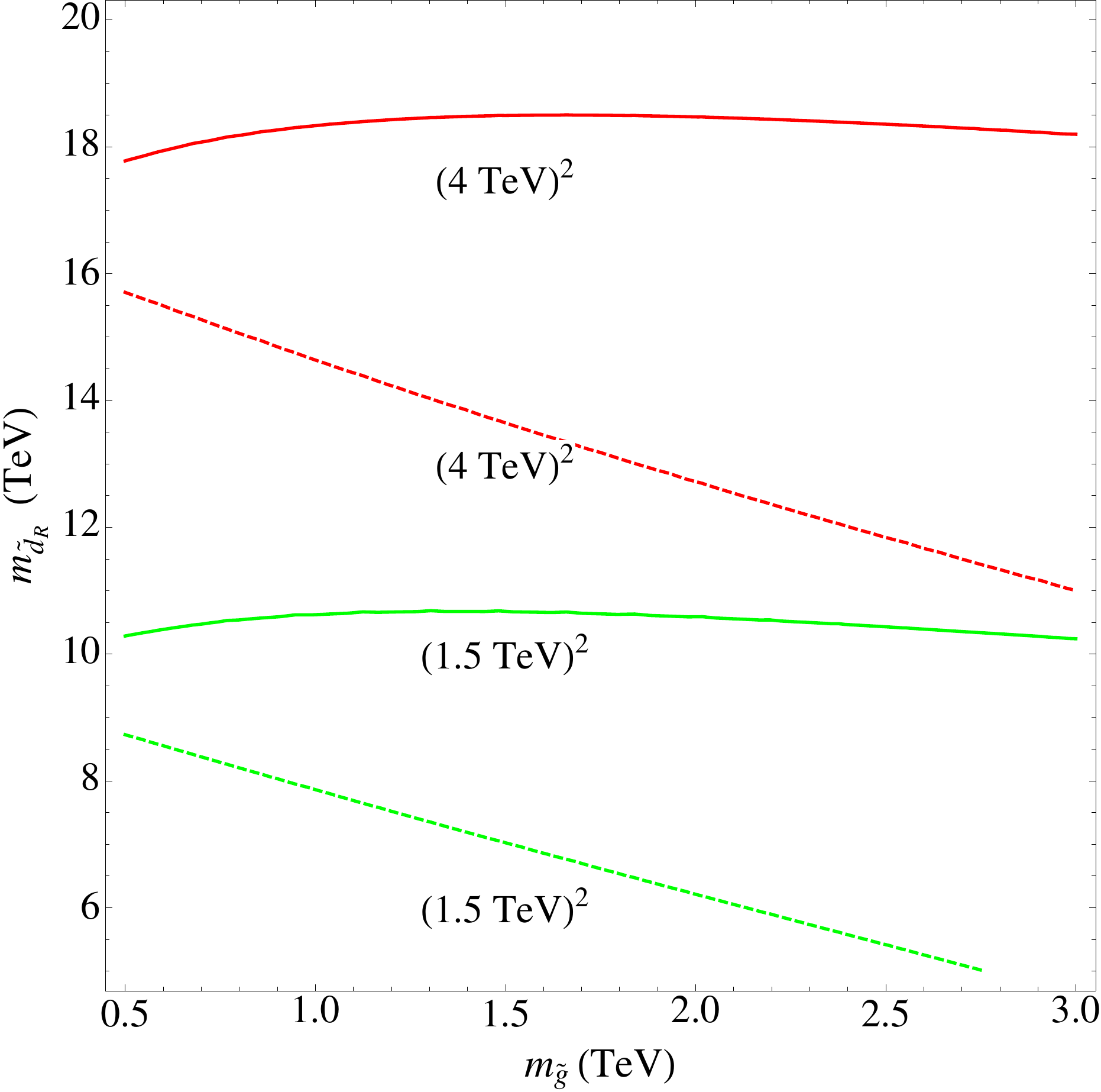}
\caption{Contour plots for the model of section \ref{SEC:DudasModel}. Along the contour, $|\Delta \epsilon_K|=\Delta \epsilon_K(exp)$. The dashed lines correspond to exactly Dirac gauginos, while the solid lines are purely Majorana, as in the original model of \cite{Dudas:2013pja}. In the left plot, the left-handed sbottom mass is set equal to that of the gluino; in the right plot, the left-handed sbottom is fixed at 1 TeV.  
The two lines correspond to $m_{\tilde{d}_R}^2-m_{\tilde{b}_R}^2= (1.5,4 \ {\rm TeV})^2$. The remaining parameters are chosen as $|V_{23}^d|=0.04,\sin(\alpha_{12})=0.5$ and
$s_d^2=0.2$. }
\label{FIG:DUDASMODELPLOT}
\end{center}
\end{figure}

The bounds on $\Delta m_K$ are easily satisfied by this model, so we focus directly on the bounds on $\epsilon_K$. We obtain, allowing $C_{\epsilon_K} \in [0.66,1.73]$ at $99\%$ confidence level:
\begin{align}
\frac{|\Delta \epsilon_K|}{|\epsilon_K (SM)| 0.73} \simeq& \frac{4.25}{0.73\times 2.04\times 10^{-3}} \frac{1}{3} \frac{m_K f_K^2}{m_1^2 \sqrt{2} \Delta m_K (\mathrm{exp})} |V_{23}^d|^2 \frac{m_d}{m_s}s_d^2 \sin 2 \tilde{\alpha}_{12} \frac{|m_{\tilde{d}_R}^2 - m_{\tilde{b}_R}^2|}{m_1^2} \tilde{f}_5 \nn\\
=& 0.07 \times \left( \frac{|V_{23}^2|}{0.04}\right)^2 \left(\frac{s_d^2}{0.2}\right) \left(\frac{\sin 2\tilde{\alpha}_{12}}{\sqrt{3}/2}\right) \left( \frac{ |m_{\tilde{d}_R}^2 - m_{\tilde{b}_R}^2|}{1\ \mathrm{TeV}^2} \right) \left( \frac{10\ \mathrm{TeV}}{m_1^2} \right)^2 \left(\frac{\tilde{f}_5}{1.1}\right)
\end{align}
where we have used $m_D = 2\ \mathrm{TeV}$, $m_{L3}=m_{\tilde{b}_{L}} = 1\ \mathrm{TeV}$, $m_1 = 10\ \mathrm{TeV}$ to evaluate $\tilde{f}_5$.  The results from \cite{Dudas:2013pja} are compatible with the $95\%$ confidence level bound $C_{\epsilon_K} \in [0.77,1.41] $ and we show the comparison in figure \ref{FIG:DUDASMODELPLOT}.

In this model, the Dirac gluino offers an improvement by roughly a factor of four over the Majorana case. Again, this is in agreement with sec. \ref{SEC:argument} since the dominant contribution comes from $C_4$ where the chirality-flip process adds to the same-chirality one instead of cancelling it.

\section{A Diversion: how to fake a gluino}
\label{SEC:FakeGaugino}

We saw previously that large suppression of FCNC and production of coloured particles
can be obtained in two different ways: 
\begin{itemize}
\item Large Dirac mass $m_D \gg M,M_{\chi}$, due to the underlying R-symmetry, in the
mass insertion approximation.
\item Large Majorana mass $M \gg m_D,M_{\chi}$, due to small couplings of the light ``fake gaugino'' fermion to quarks/squarks. 
\end{itemize}

The second case can be realised in two distinct ways. \\
i) We can have a scenario with a very moderate hierarchy and without a see-saw mass: we can take for example $M_\chi \sim$ TeV, $M \sim 10$ TeV, $1 \ {\rm TeV} \lesssim m_{\tilde{q}} \lesssim 5$ TeV and an arbitrarily small Dirac mass. In particular, we need only consider the gluino  as being so heavy (the other gauginos could be somewhat lighter). In this case, all of the masses would be generated by F-term supersymmetry-breaking. The Dirac mass is then automatically suppressed, as can be checked
by writing explicitly the Dirac mass term with the help of a chiral spurion superfield. Alternatively, there can be also a small D-term which would explain the smallness of the Dirac mass. Here $R_{12}^{\tilde{g}} \sim m_D/M$, so the mixing between the gauginos and the fake gaugino could be almost arbitrarily small. \\
ii) A second way is by having a large, intermediate scale gluino mass. A theoretical motivation for this case is gauge coupling unification. According to
\cite{bachas}, MSSM with additional adjoint chiral fields leads to a good unification of couplings at the string scale for adjoint masses around $10^{12}$ GeV. In the case they
considered, the low-energy effective theory is just the MSSM. From the gauge unification viewpoint however, we can switch the masses of the gauginos/gluinos with those of the chiral adjoint fermions, keeping the scalar adjoint masses  heavy.  This switch will not affect gauge coupling unification at one-loop, whereas it will significantly change phenomenology, which we call ``fake split SUSY'' for obvious reasons in what follows.
In this section we therefore consider in more detail this scenario and comment on its qualitative phenomenological consequences. 

Case i) is clearly easy to justify. Before discussing phenomenological implications, let elaborate more about case ii). The obvious question is the stability of the hierarchy $M \gg m_D,M_{\chi}$ under radiative corrections. For this, we need to consider the effective theory when we integrate out the gauginos and the sfermions. The adjoint fermion $\chi$ (the ``fake gaugino'') has no tree-level renormalisable couplings to the squarks and sleptons, but it does couple via the gauge current to the gaugino $\lambda$ and the adjoint scalar $\Sigma$: the relevant terms are
\begin{align}
\mathcal{L} \supset& - \bigg( \frac{M}{2} \lambda^a \lambda^a + \frac{M_{\chi}}{2} \chi^a \chi^a + \frac{1}{2} B_{\Sigma} \Sigma^a \Sigma^a  +  i \sqrt{2} g  f^{abc} \ov{\Sigma}^a \lambda^b \chi^c  + h.c. \bigg)  - m_{\Sigma}^2 \Sigma^a \ov{\Sigma}^a  \nn\\ & - ( m_D \Sigma^a + \ov{m_D} \ov{\Sigma}^a )^2 - ( m_D \lambda^a \chi^a + c.c.) . 
\end{align}
On the second line we included the terms coming from the Dirac gaugino mass term, which necessarily also generates the term $( m_D \Sigma^a + \ov{m_D} \ov{\Sigma}^a )^2 $. We do not absorb these into $m_{\Sigma}, B_{\Sigma}$ because these corrections are RGE invariant and therefore apply at any renormalisation scale \cite{Jack:1999ud,Jack:1999fa,Goodsell:2012fm}. Instead we define
\begin{align}
\hat{B}_\Sigma \equiv B_\Sigma + 2 m_D^2 \ , \
\hat{m}^2_\Sigma \equiv  m_\Sigma^2 + 2 |m_D|^2 . \label{hats}
\end{align}
Since we are making the logical assumption that the adjoint scalars are at least as massive as the other scalars in the theory, we can integrate them out along with the gaugino $\lambda$: at one loop we generate a term $M_\chi$ of
\begin{align}
M_\chi = 2g^2 C_2 (G) \hat{B}_\Sigma M \int \frac{d^4 p}{(2\pi)^4} \frac{p^2}{((p^2 + m_D^2)^2 + M^2 p^2)((p^2 + \hat{m}_\Sigma^2)^2 - \hat{B}_\Sigma^2)} \ , 
\end{align} 
which gives to leading order in $B_\Sigma/m_\Sigma^2, m_D/M$
\begin{align}
M_\chi =& \frac{2 C_2 (G)g^2}{16\pi^2} \times \left\{ 
\begin{array}{cl} 
\frac{\hat{B}_\Sigma }{M} \bigg( 1 - \log \frac{M^2}{\hat{m}_\Sigma^2} \bigg)  & M \gg \hat{m}_\Sigma \ ,  
\\  \frac{\hat{B}_\Sigma }{\hat{m}_\Sigma^2} M  & \hat{m}_\Sigma \gtrsim M .
\end{array} \right. 
\end{align}
This clearly prevents an arbitrary hierarchy between $M$ and $M_\chi$. We might consider simply ignoring $\hat{B}_\Sigma$; however, it will always have a D-term contribution from the Dirac mass, so that without tuning we can say $|\hat{B}_\Sigma| \gtrsim |m_D|^2$. More honestly, we should look if there can be a symmetry preventing the generation of such a term. Indeed this is the case: If we rotate the adjoint field $\Sigma$ then this prevents both $M_\chi$ and $B_\Sigma$, but also prevents the Dirac mass $m_D$. However, if we break this symmetry with the vev of a field $\phi$ such that $\phi/M_{\mathrm{high}} \equiv \epsilon $ then we generate 
\begin{align}
m_D &\sim \epsilon M \quad , \quad 
M_{\chi} \sim  \epsilon^2 M \quad , \quad
B_{\Sigma }  \sim \epsilon^2 M^2 \sim m_D^2 
\end{align}
and thus the above contribution is irrelevant: the see-saw (and direct) masses for the ``fake'' gluino are of order $m_D^2/M$ where the scale is controlled by the parameter $\epsilon$. We also note that since this hierarchy is protected by the approximate symmetry, it is not affected by renormalisation group running from above the SUSY-breaking scale\footnote{ In terms of the parameter ${\hat B}_{\Sigma}$ defined in (\ref{hats}), we find $ \delta {\hat B}_{\Sigma} \sim  (M M_\chi -m_D^2) 
\frac{g^2}{16 \pi^2} \log\left(\frac{\Lambda}{M}\right) $.}
\bea\label{hier}
&&\delta M_{\chi}\sim \epsilon^2 \frac{g_s^2}{16 \pi^2} M\,, \\ \nonumber
&&\delta B_{\Sigma}\sim M M_\chi \frac{g^2}{16 \pi^2} \log\left(\frac{\Lambda}{M}\right) \sim \epsilon^2\frac{g^2}{16 \pi^2} M^2 \log\left(\frac{\Lambda}{M}\right).
\eea

Taking  $M \sim m_{\tilde q} \sim m_{\Sigma} \sim 10^{12}$ GeV and assuming that the ``fake'' gluino mass is of order $M_{\chi}\sim 1$ TeV, this fixes the parameter $\epsilon$ to be of order $10^{-4}$ (so that we could take $\bra \phi \ket \sim M, M_{\mathrm{high}} \sim M_{\mathrm{GUT}}$). Hence we get the following masses
\begin{equation}
 M\sim 10^{12}\text{GeV}\gtrsim m_{\Sigma}~~\gg~~m_D,~\sqrt{B_{\Sigma}} \sim 10^{8}\text{GeV}~~\gg~~~M_{\chi}\sim 1\text{TeV~.} \label{w2}
\end{equation} 

If the switch of masses is also performed for the wino/bino $\leftrightarrow$ fake wino/bino, the resulting low-energy effective theory in this case is different compared
to standard split SUSY. Indeed, we should consider whether there are any light higgsinos remaining in the spectrum: in split SUSY, there is an R-symmetry that protects the mass of the higgsinos, whereas we have broken this, and we would expect the higgsinos to obtain a mass through diagrams similar to the one considered above:
\begin{equation}
\mu \simeq \frac{1}{4} 2g^2_Y  M B_\mu \tilde{I}_4 (m_h^2, m_H^2, M_{\tilde{B}^1}^2, M_{\tilde{B}^2}^2) + \frac{3}{4} 2g^2_2  M B_\mu \tilde{I}_4 (m_h^2, m_H^2, M_{\tilde{W}^1}^2, M_{\tilde{W}^2}^2)  \ , \label{higgino1}
\end{equation} 
where $M_{\tilde{B}^i},M_{\tilde{W}^i} $ with $i=1,2$ are the masses for the bino and wino eigenstates respectively (before electroweak symmetry breaking) and $m_h$ ($m_H$) are the
light (heavy) mass parameters in the Higgs sector,
\begin{equation}
m_h^2 \simeq \frac{m_{h_u}^2 m_{h_d}^2 - B_{\mu}^2}{m_{h_u}^2+m_{h_d}^2} \quad ,
\quad m_H^2 \simeq m_{h_u}^2+m_{h_d}^2 \ . \label{higgino2}
\end{equation}
In writing (\ref{higgino1}) we neglected
$M_{\chi}$ in the loop. In this case, a more compact form for the integrals is, for example
\begin{equation}
\tilde{I}_4 (m_h^2, m_H^2, M_{\tilde{W}^1}^2, M_{\tilde{W}^2}^2) = \int \frac{d^4 p}{(2 \pi)^4} \frac{p^2}{(p^2+m_h^2)(p^2+m_H^2) [(p^2+m_D^2)^2+M_{\tilde{W}}^2 p^2]} . \label{higgino3}
\end{equation}  
Whereas the general expression is rather involved, in the limit $M \gg m_H $ and (for simplicity) with equal gaugino mass parameters for $SU(2)$ and $U(1)$ factors 
$M_{\tilde{W}} = M_{\tilde{B}} \simeq M $, it simplifies
to
\begin{equation}
\mu \simeq \frac{g_Y^2 + 3 g_2^2}{32\pi^2} \frac{B_\mu}{M - m_H^2/M} \log \frac{m_H^2}{M^2} \ . \label{higgino4}
\end{equation}
However, this can be repaired in a similar fashion: we can suppose that the Higgs fields are charged under the same $U(1)$ symmetry that the adjoints are charged under. This would suppress the $\mu$ and $B_\mu$ terms, and also prevent any superpotential couplings between the adjoints and the higgsinos. We would have $\mu \sim \epsilon^2 M, B_\mu \sim \epsilon^2 M^2 $ so we would have $B_\mu \gg |\mu|^2$ and the heavy Higgs scalars would be parametrically heavier than the electroweak scale. In this scenario we effectively take infinite $\tan \beta$ and require the down-quark and lepton Yukawa couplings non-holomorphic and generated in the high-energy theory (see e.g. \cite{Ibanez:1982xg,Davies:2011mp}).\footnote{There is another solution, where instead we extend the Higgs sector by another pair of doublets. These could be consistent with unification \emph{at any scale}; this is being explored in another work \cite{DGSplit}.}

Then, in split SUSY the effective lagrangian contains higgs/higgsino/gaugino
couplings
\begin{align}
& {\cal L}_{\rm eff} \supset  -\frac{H^{\dagger}}{\sqrt{2}} ({\tilde g}_u \sigma^a {\tilde W}^a + {\tilde g'}_u {\tilde B}) \ {\tilde H}_u -  \frac{H^{T} \epsilon}{\sqrt{2}} (- {\tilde g}_d \sigma^a {\tilde W}^a + {\tilde g'}_d {\tilde B}) \ {\tilde H}_d  \label{w5}.
\end{align}
In usual split SUSY, ${\tilde g}_u = g \sin \beta , {\tilde g}_d = g \cos \beta ,  {\tilde g'}_u = g' \sin \beta , {\tilde g'}_d = g' \cos \beta$; however, in our case these couplings will be strongly suppressed by the fake gaugino/bino compositions $R_{12}$, $R'_{12}$. If the adjoint superpotential couplings $W \supset \lambda_S H_d S H_u + 2 \lambda_T H_d T H_u$ had not been suppressed, then they would have provided couplings of the same form. Instead, the absence of such couplings at low-energy could be therefore a signature of a remote $N=2$ supersymmetric sector, instead of a more
conventional split SUSY spectrum.  

Finally, in the absence of couplings $\lambda_{S,T}$, the model has difficulties to
accommodate a good dark matter candidate, due to the small couplings of the fake electroweakinos to quarks and leptons.

\subsection{Phenomenological consequences}

In the context of split SUSY, where squarks are very heavy compared to the gluino, one striking experimental signature is the long lifetime of the gluino and associated displaced vertices or (for even heavier squarks) gluino stability. Indeed the lifetime of the gluino could be sufficiently long to propagate on macroscopic distances in detectors \cite{Toharia:2005gm, Hewett:2004nw, Gambino:2005eh}. This lifetime, in the standard split SUSY context, can be estimated in an approximate manner according to \cite{Gambino:2005eh} as follows
\begin{equation}
 \tau_{\tilde g}= \frac{4 \ \text{sec}}{N}\times\left(\frac{m_{\tilde q}}{10^9\text{GeV}}\right)^4\times\left(\frac{1 \ \text{TeV}}{M}\right)^5\,, \label{w3}
\end{equation} 
where $N$ is a quantity varying with $M$ and $m_{\tilde q}$ but of order one for our 
range of masses.

As we saw in the previous sections, the fake gluino couplings are altered by the diagonalisation of the gluino mass matrix and contain a tiny contribution of the original gluino gauge coupling, proportional to $R_{12}^{\tilde{g}} \sim m_D/M$. In case
i) above, the mixing between the gauginos and the fake gaugino could be almost arbitrarily small by having $m_D \ll\ $TeV, meaning that the fake gluino could still have displaced vertices without requiring large mass scales. Particularly interesting is the case where
the usual gluinos are not accessible (they are heavier than say 5 TeV), whereas some
of the squarks are. Displaced vertices /long lifetime for the fake gluino with light
squarks would be a direct probe of a high-energy $N=2$ supersymmetric spectrum. 
Pair production of  faked gluinos in this case lead to displaced vertices, since although some squarks could be light, their small couplings to the fake gluino suppresses such processes. On the
other hand, direct squark production is possible, but subsequent squark decays to quarks/neutralinos go dominantly through the Higgsino components and corresponding Yukawas couplings. They are therefore unsuppressed only for third generation squarks ( and eventually third generation sleptons if similar arguments are applied to the other gauginos).  Of course, the heavier the usual gluino, the bigger the fine-tuning needed
in order to keep a squark to be light. Some fine-tuning, moderate for gluino mass below $10$ TeV or so, is unavoidable for such a scenario to be realised in nature. However, its
very different phenomenological implications could be worth further study.  

In case ii) above, the fake gluino couplings to quarks/squarks are proportional to
$g_s R_{12} \simeq g_s \frac{m_D}{M} \sim \epsilon $  and encodes the small gluino composition of the lightest fermion octet. According to our numerical choice of masses we get $R_{12}\sim \epsilon \sim 10^{-4}$. This affects therefore the fake gluino lifetime, which has to be modified according to 
\begin{align}
  \tau_{\chi}=& \frac{4 \times 10^{28} \ \text{sec}}{N} \times \left(\frac{10^{-4}}{R_{12}^{\tilde{g}}} \right)^2 \left(\frac{10^{-4}}{R_{12}^{\chi^0}} \right)^2  \times\left(\frac{m_{\tilde q}}{10^{12} \text{GeV}}\right)^4 \times\left(\frac{1 \ \text{TeV}}{M_{\chi}}\right)^5 \sim 10^{21} \ \text{years}
\label{w4}
\end{align} 
where we define $R^{\tilde{g}}$ and $R_{12}^{\chi^0} $ to be the rotation matrices for the gluino and neutralino respectively. For the scales given, this lifetime is hence longer than the age of the universe, and so we should make sure that fake gluinos are not produced in the early universe\footnote{For more discussion of this issue we refer the reader to \cite{DGSplit}}. 

We could also consider different moderate hierarchies with
interesting low-energy implications. For example, let us suppose that $M_{\chi} \sim m_D \sim \mathrm{TeV}$ and
gluino and squark masses $M \sim m_{\tilde q} \sim 100$ TeV, while the higgsinos remain light; in split SUSY gluino decays are prompt inside the detector, but in our ``fake split SUSY'' case, now $R_{12}^{\tilde{g}} \sim 10^{-2}$ and we can take $R_{12}^{\chi^0} \sim 1$. The gluino propagation length is increased by a factor of $10^{4} $ and the vertex
starts to become displaced. Although the squarks are still very heavy, they could
produce testable CP violating FCNC effects in the Kaon system ($\epsilon_K$).

\section{Conclusions}
\label{SEC:Conclusions}

Flavour physics sets severe constraints on supersymmetric models of flavour. In models
in which the scale of mediation of supersymmetry breaking is similar or higher than
the scale of flavour symmetry breaking, fermion masses and mixing hierarchies are
correlated with the flavour structure of superpartners. In the MSSM constructing
a fully successful flavour model of this type is difficult and usually requires the
simultaneous presence of several ingredients like abelian and non-abelian symmetries.
At first sight, flavour models with Dirac gauginos are simpler to build, due to the 
flavour suppression argued in the literature in their R-symmetric pure Dirac limit,
for gluinos heavier than squarks. 
In this paper, we found that this suppression is only strong in the near-degeneracy
(mass insertion approximation) limit, whereas in most flavour models this approximation
is not valid. 

We analysed the simplest Dirac flavour models with abelian symmetries realising various
degrees of alignment of fermion and scalar mass matrices and for non-abelian symmetries
realising a natural supersymmetric spectrum with heavy first two generations. 
We found only a moderate improvement in the flavour constraints over the 
MSSM case. We also showed in an explicit example in section \ref{SEC:ALIGNMENT} that due to cancellations
in the Majorana  case, it is even possible that a Dirac model is for some parts of the parameter space more constrained
than its MSSM cousin.

We also considered generalised Lagrangians with both Majorana and Dirac masses, by not
imposing an R-symmetry in the UV. We considered, in particular, the case in which the
gluino Majorana mass is very large compared to that of the chiral octet fermion and the
Dirac mass $M \gg M_{\chi},m_D$. This led to the scenario dubbed ``fake gluino'' in which the light adjoint fermions are not the $N=1$ partners of the gauge fields, but the other fermions in the  $N=2$ gauge multiplets. In this case, couplings of the light ``fake gluino'' 
to quarks are suppressed parametrically by the ratio $m_D/M$. This leads to a potentially new exotic phenomenology
in which squarks can be light and accessible experimentally, while the light adjoint fermions can be long-lived and generate displaced vertices or escape detection. Experimental discovery of a squark and simultaneously of long-lived light gluinos would be spectacular evidence of such a spectrum. 
An extreme case with heavy gluinos and light adjoint fermions is obtained by pushing a
Majorana gluino mass and squark masses to an intermediate scale $M \sim 10^{12}$ GeV,
which leads to good gauge coupling unification. The outcome is similar in spirit to split supersymmetry, with however light adjoint fermion couplings to quarks and (for electroweakinos) to higgs/higgsino which are highly suppressed compared to split supersymmetry.


\section*{Acknowledgements}  
We would like to thank Robert Ziegler for interesting discussions. M.~D.~G.~ would like to thank Karim Benakli, Luc Darm\'e and Pietro Slavich for useful discussions and collaboration on the related work \cite{DGSplit}.  We thank the Galileo Galilei Institute for Theoretical Physics for the hospitality and the INFN for partial support in the initial stage of this work. This  work was supported in part by the European ERC Advanced Grant 226371 MassTeV, the French ANR TAPDMS ANR-09-JCJC-0146, the contract PITN-GA-2009-237920 UNILHC and the Marie-Curie contract no. PIEF-GA-2012-330060 BMM@3F. P.~T. is supported in part by Vrije Universiteit Brussel through the Strategic Research Program ``High-Energy Physics'', in part by the Belgian Federal Science Policy Office through the Inter-university Attraction Pole P7/37 and in part by FWO-Vlaanderen through project G011410N.

\appendix

\section{K and B meson mixing in supersymmetry}\label{MSSMbox}

\subsection{From the Lagrangian to Feynman rules}\label{feynrules}

In MSSM, the dominant contribution to K and B meson mixing comes from a box diagram with squarks and gluinos propagating in the loop. Starting from the superfield Lagrangian, we have
\bea\label{qsqgl}
\cL_{MSSM}&\supset&\int d^4\theta\ Q^\dag e^{2g_sV^aT^a}Q+\ov{D}^\dag e^{-2g_sV^aT^{a*}}\ov{D}\nonumber
\\
&\supset& -\sqrt{2}g_s\Big[ \tilde{d}_{Lxi}^*T^a_{xy}\lambda^{a\alpha}  d_{Lyi\alpha} -\tilde{d}_{Rxi}T_{xy}^{a*}\lambda^{a\alpha}d_{Ryi\alpha}^c\Big]+h.c.\,,
\eea
where $g_s$ is the strong coupling constant, $i=1,2,3$ is the flavour index, $T^a_{xy}$ are the $SU(3)$ generators and $\lambda^{a\alpha}$ is the gluino Weyl fermion. Also, the fermion in the chiral superfield $\ov{D}$ is denoted by $d_{Rxi\alpha}^c=(d_{Rxi})^c_\alpha$ and describes the charge conjugate of the right-handed down quark field. Its scalar superpartner is $\tilde{d}_{Rxi}^*$.

After adopting four-component notation
\be
d=\left(\begin{array}{c}d_{L\alpha} \\\ov{d^c_{R}}^{\dot{\alpha}}\end{array}\right)\,;\quad \tilde{g}^a=\left(\begin{array}{c} \lambda_\alpha^a \\ \ov{\lambda}^{a\,\dot{\alpha}} \end{array}\right)\,,
\ee
and using identities
\be
\ov{\Psi^c}_i \Gamma_I \Psi_j^c=-(-1)^Ag_{IJ}\ov{\Psi}_j \Gamma_J \Psi_i
\ee
where $\Gamma_I=\{\mathbb{I},\,\gamma^5,\,\gamma^\mu\gamma^5,\,\gamma^\mu,\,\Sigma^{\mu\nu},\,\Sigma^{\mu\nu}\gamma^5 \}$, $g_{IJ}=\mathrm{diag}(1,\,1,\,1,\,-1,\,-1,\,-1)$ and $(-1)^A=+1\,(-1)$ for a commuting (anticommuting) $\Psi_i$, (\ref{qsqgl}) can be written as
\be\label{qsqgl3}
-\sqrt{2}g_sT_{xy}^{a}\Big[ \tilde{d}_{Lxi}^*\ov{\tilde{g}}^a P_Ld_{yi} -\tilde{d}^*_{Rxi}\ov{\tilde{g}}^aP_Rd_{yi}\Big]+h.c.
\ee
This last expression can also be written using charge conjugated fields
\be \label{qsqgl4}
-\sqrt{2}g_sT_{xy}^{a}\Big[ \ov{\tilde{g}}^a P_Rd_{xi}^c\tilde{d}_{Lyi} -\ov{\tilde{g}}^aP_Ld_{xi}^c\tilde{d}_{Ryi}\Big]+h.c.\,.
\ee
Before writing down the Feynman rules for couplings (\ref{qsqgl3}) and (\ref{qsqgl4}), we switch to the squark mass eigenstate basis. 

\begin{figure*}[!t]
\includegraphics[scale=0.7]{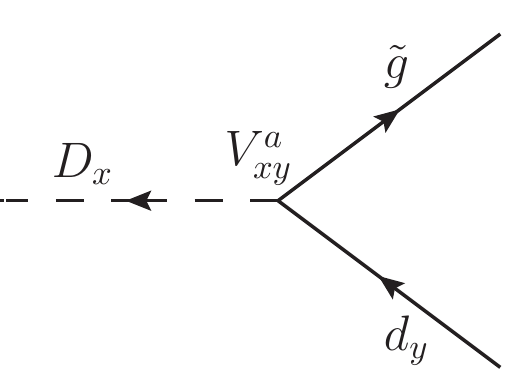}
\includegraphics[scale=0.7]{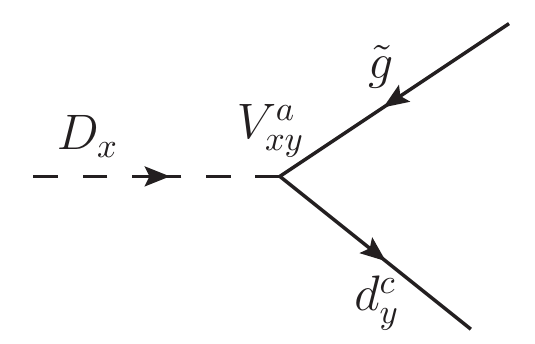}
\includegraphics[scale=0.7]{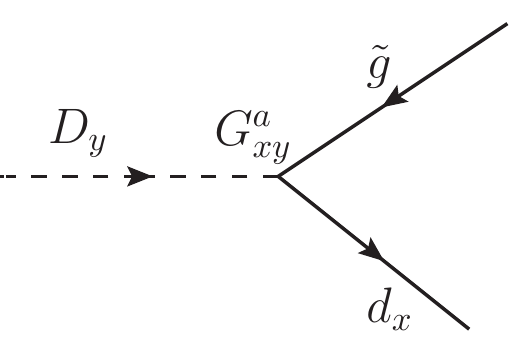}
\includegraphics[scale=0.7]{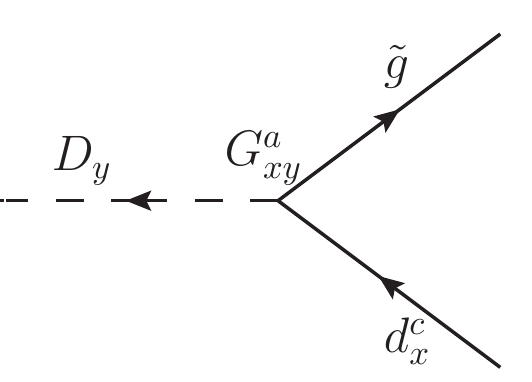}
\caption{ {\small Feynman rules for gluino - squark - quark vertices. }}
\label{Vxy}
\end{figure*}

Going first to the basis where the down quark mass matrix and the gluino - squark - quark coupling are diagonal, one can write 
\be
d_{L} \rightarrow V_Ld_L\text{~,~~} \overline d_{R}^c \rightarrow V_R \overline d_R^c\,.
\ee
The down squark mass matrix is now denoted by $m_{\tilde{d}}^2$
\be\label{msquark1}
\cL_{m_{\tilde{d}}}= - \tilde{d}^\dag\,m_{\tilde{d}}^2\,\tilde{d}\,,\quad \tilde{d}=\left(\begin{array}{c}\tilde{d}_{Li} \\\tilde{d}_{Ri}\end{array}\right)
\ee
and can be diagonalised by the unitary matrix $Z_{IJ}$
\be\label{msquark2}
\tilde{d}_{Li}=Z_{iI}\tilde{D}_I\,,\quad \tilde{d}_{Ri}=Z_{i+3I}\tilde{D}_I
\ee
such that
\be\label{masseigen}
\cL_{m_{\tilde{d}}}= - \tilde{D}^*_I\,m_{I}^2\,\tilde{D}_I\,,\quad m^{2}=Z^\dag m_{\tilde{d}}^{2}\,Z
\ee
where $\tilde{D}_I$ with $I=1,\dots,6$ is the squark mass eigenstate, $i=1,2,3$ is the flavour index and $m^2=\mathrm{diag}(m_I^2)$ is the diagonal matrix of the mass eigenstates. Then, (\ref{qsqgl3}) and (\ref{qsqgl4}) are written as (we denote $(W^\dag)_{IJ}\equiv Z^\dag_{IJ}$)
\be
-\sqrt{2}g_sT_{xy}^{a}\Big[ \tilde{D}_{Ix}^*W^\dag_{Ii}\ov{\tilde{g}}^a P_Ld_{yi} -\tilde{D}_{Ix}^*W^\dag_{Ii+3}\ov{\tilde{g}}^aP_Rd_{yi}\Big]+h.c.
\ee
and
\be
-\sqrt{2}g_sT_{xy}^{a}\Big[ \ov{\tilde{g}}^a P_Rd_{xi}^cW_{iI}\tilde{D}_{Iy} -\ov{\tilde{g}}^aP_Ld_{xi}^cW_{i+3I}\tilde{D}_{Iy}\Big]+h.c.\,,
\ee
where W is defined by
\be 
W=\begin{pmatrix}
V_L^{\dagger} Z_{LL} & V_L^{\dagger} Z_{LR} \\
V_R^{\dagger} Z_{RL} & V_R^{\dagger} Z_{RR}
\end{pmatrix}\,.
\ee
The corresponding Feynman rules for the vertices are (see figure \ref{Vxy})
\be
V^a_{xy}=-i\sqrt{2}g_sT_{xy}^a(W^\dag_{Ii}P_L-W^\dag_{Ii+3}P_R)\,,\quad G^a_{xy}=-i\sqrt{2}g_sT_{xy}^a(W_{iI}P_R-W_{i+3I}P_L)
\ee

\subsection{From the amplitude to the effective action}

\begin{figure*}[!t]
\includegraphics[scale=0.7]{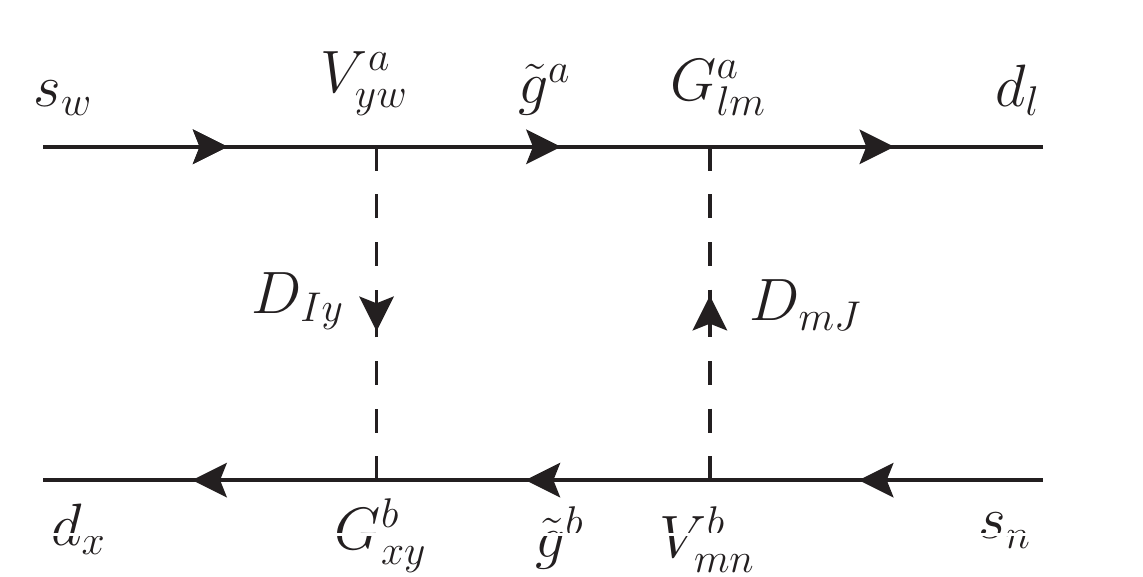}
\includegraphics[scale=0.7]{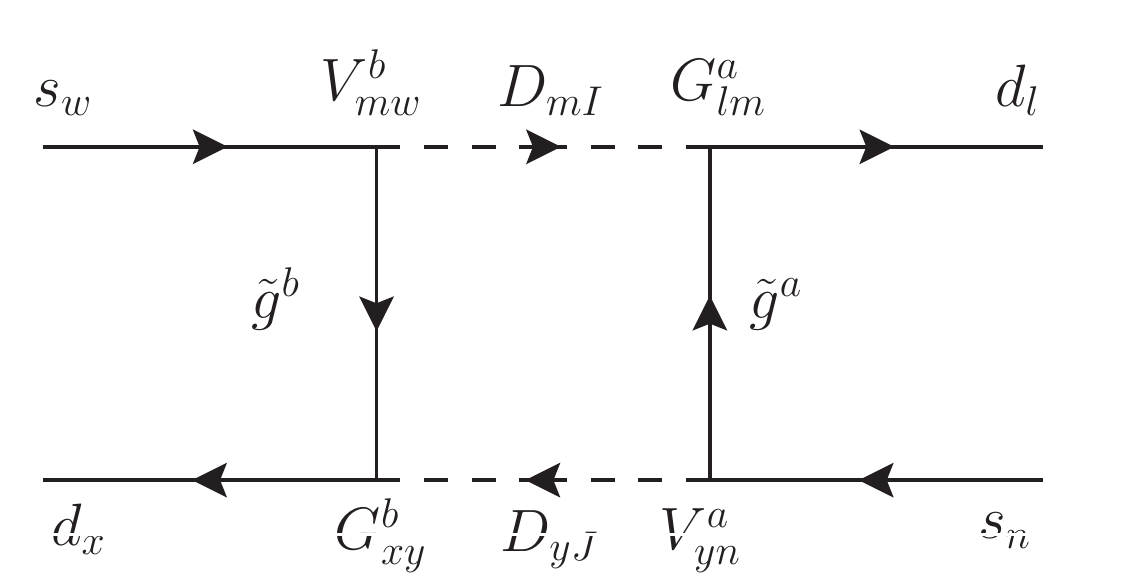}
\includegraphics[scale=0.7]{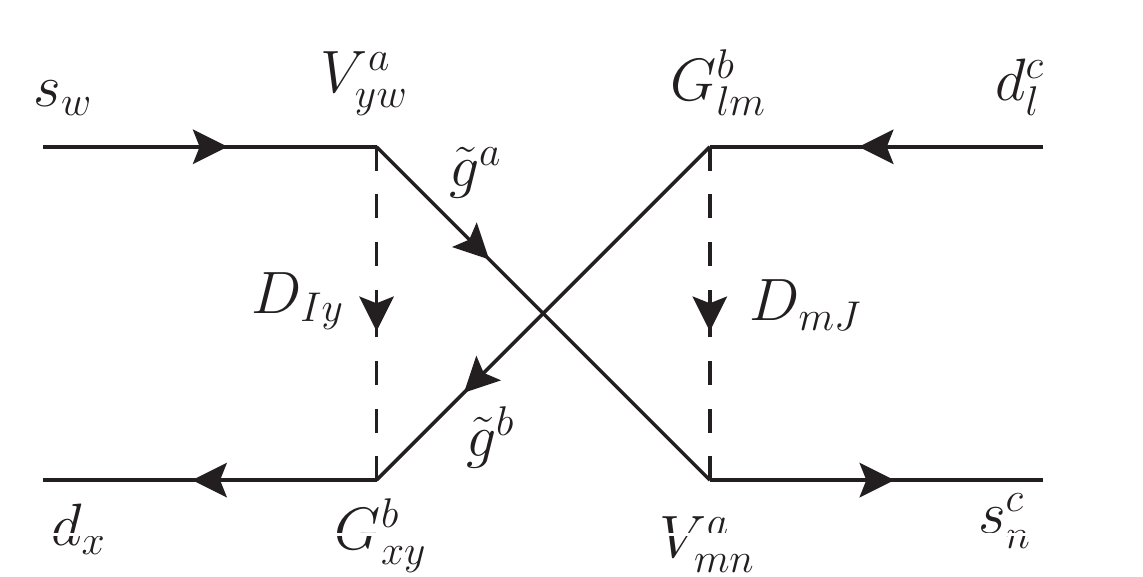}
\includegraphics[scale=0.7]{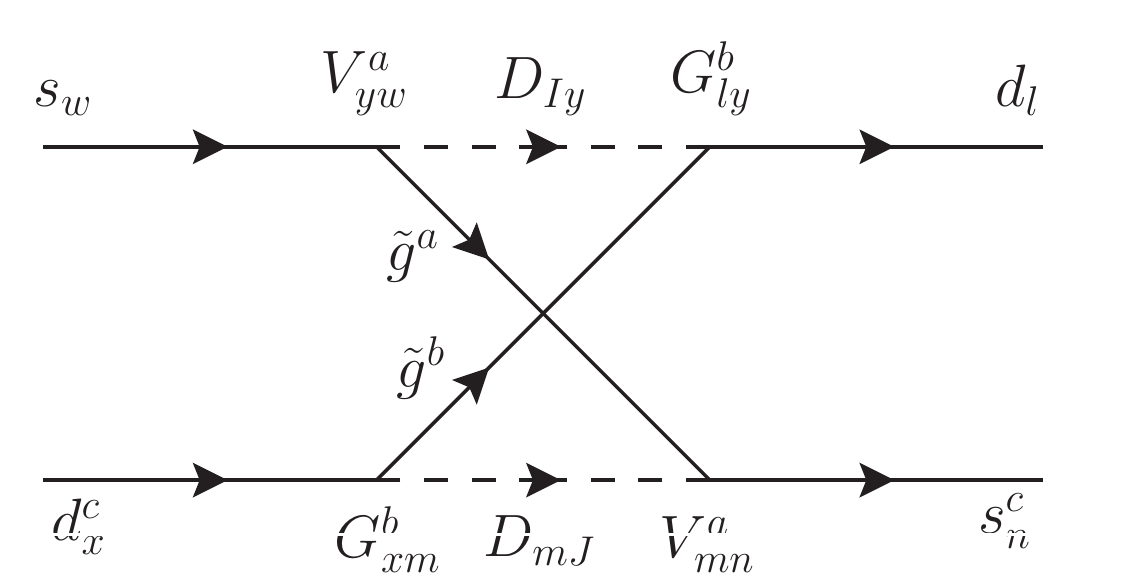}
\caption{ \small The four box diagrams (denoted $M_{1,2,3,4}$ from top left to bottom right) that contribute to $K$-$\ov{K}$ mixing. The fields $s_w$, $d_x$, $d_l$ and $s_n$ have 4-momenta $p_{1,2,3,4}$.
}
\label{boxdiagrams}
\end{figure*}

The amplitudes of the diagrams in figure \ref{boxdiagrams} are (we neglect external momenta) \cite{hagelin}
\bea
&&iM_1=-\int\!\! {d^4p\over (2\pi)^4}{i\over p^2-m_{I}^2}{i\over p^2-m_{J}^2} \ov{d}_x G_{xy}^b {i(\hcancel{p}+M_{\tilde{g}})\over p^2-M^2_{\tilde{g}}}V_{mn}^bs_n\ov{d}_l G_{lm}^a {i(\hcancel{p}+M_{\tilde{g}})\over p^2-M^2_{\tilde{g}}}V_{yw}^as_w\nonumber
\\
&&iM_2=\int\!\! {d^4p\over (2\pi)^4}{i\over p^2-m_{I}^2}{i\over p^2-m_{J}^2} \ov{d}_x G_{xy}^b {i(\hcancel{p}+M_{\tilde{g}})\over p^2-M^2_{\tilde{g}}}V_{mw}^bs_w\ov{d}_l G_{lm}^a {i(\hcancel{p}+M_{\tilde{g}})\over p^2-M^2_{\tilde{g}}}V_{yn}^as_n\nonumber
\\
&&iM_3=\int\!\! {d^4p\over (2\pi)^4}{i\over p^2-m_{I}^2}{i\over p^2-m_{J}^2} \ov{d}_x G_{xy}^b {i(\hcancel{p}+M_{\tilde{g}})\over p^2-M^2_{\tilde{g}}}G_{lm}^bd_l^c\ov{s^c_n} V_{mn}^a {i(\hcancel{p}+M_{\tilde{g}})\over p^2-M^2_{\tilde{g}}}V_{yw}^as_w\nonumber
\\
&&iM_4=-\int\!\! {d^4p\over (2\pi)^4}{i\over p^2-m_{I}^2}{i\over p^2-m_{J}^2} \ov{d}_l G_{ly}^b {i(\hcancel{p}+M_{\tilde{g}})\over p^2-M^2_{\tilde{g}}}G_{xm}^bd_x^c\ov{s^c_n} V_{mn}^a {i(\hcancel{p}+M_{\tilde{g}})\over p^2-M^2_{\tilde{g}}}V_{yw}^as_w\nonumber
\eea
where $d_x$, $s_w$ etc. now denote commuting spinors. The total amplitude is simplified by using $SU(3)$ generator identities
\be
T_{xy}^aT_{mn}^aT_{lm}^bT_{yw}^b={1\over 36}(\delta_{xw}\delta_{nl}+21\delta_{xn}\delta_{lw})\,;\ T_{xy}^aT_{nm}^aT_{ml}^bT_{yw}^b={1\over 36}(10\delta_{xw}\delta_{nl}-6\delta_{xl}\delta_{nw})
\ee
as well as the Fierz identities such as
\bea
&&\ov{\Psi}_1P_L\Psi_2\ov{\Psi}_3P_R\Psi_4={1\over 2}\ov{\Psi}_1P_L\gamma^\mu\Psi_4\ov{\Psi}_3P_R\gamma_\mu\Psi_2\nonumber
\\
&&\ov{\Psi}_1P_L\gamma^\mu\Psi_2\ov{\Psi}_3P_L\gamma_\mu\Psi_4=-\ov{\Psi}_1P_L\gamma^\mu\Psi_4\ov{\Psi}_3P_L\gamma_\mu\Psi_2\quad \textrm{(same\ for\ $P_R$)}\nonumber
\\
&&\ov{\Psi}_1P_R\Psi_2\ov{\Psi}_3P_R\Psi_4={1\over 2}\ov{\Psi}_1P_R\Psi_4\ov{\Psi}_3P_R\Psi_2-{1\over 8}\ov{\Psi}_1\Sigma^{\mu\nu}P_R\Psi_4\ov{\Psi}_3\Sigma_{\mu\nu}\Psi_2\,.
\eea
We can identify an effective Lagrangian that delivers this total amplitude. In our case we use \cite{masiero}
\bea\label{Leff}
{i\over g_s^4}\cL_{eff}^{K\ov{K}}&=&W_{1K}W_{1L}\left( {11\over 36}\tilde{I}_4+{1\over 9}M_{\tilde{g}}^2I_4 \right)W_{K2}^\dag W_{L2}^\dag \ov{d}_x\gamma^\mu P_L s_x\,\ov{d}_n\gamma_\mu P_L s_n\nonumber
\\
&+&W_{4K}W_{4L}\left( {11\over 36}\tilde{I}_4+{1\over 9}M_{\tilde{g}}^2I_4 \right)W_{K5}^\dag W_{L5}^\dag\ov{d}_x\gamma^\mu P_R s_x\,\ov{d}_n\gamma_\mu P_R s_n\nonumber
\\
&+&W_{1K}W_{4L}\!\left(\!{7M_{\tilde{g}}^2I_4-\tilde{I}_4\over 3}\ov{d}_xP_L s_x\,\ov{d}_nP_R s_n\!+\!{M_{\tilde{g}}^2I_4\!+\!5\tilde{I}_4\over 9}\ov{d}_xP_L s_n\,\ov{d}_nP_R s_x\! \right)\!W_{K2}^\dag W_{L5}^\dag\nonumber
\\
&+&M_{\tilde{g}}^2W_{1K}W_{1L}I_4W_{K5}^\dag W_{L5}^\dag\left({17\over 18}\ov{d}_xP_R s_x\,\ov{d}_nP_R s_n-{1\over 6}\ov{d}_xP_R s_n\,\ov{d}_nP_R s_x \right)\nonumber
\\
&+&M_{\tilde{g}}^2W_{4K}W_{4L}I_4W_{K2}^\dag W_{L2}^\dag\left({17\over 18}\ov{d}_xP_L s_x\,\ov{d}_nP_L s_n-{1\over 6}\ov{d}_xP_L s_n\,\ov{d}_nP_L s_x \right)\nonumber
\\
&+&W_{1K}W_{4L}\tilde{I}_4W_{K5}^\dag W_{L2}^\dag\left(-{11\over 18}\ov{d}_xP_L s_x\,\ov{d}_nP_R s_n-{5\over 6}\ov{d}_xP_L s_n\,\ov{d}_nP_R s_x \right)\,.
\eea

\subsection{Loop Integrals}\label{defI}

The following loop functions are being used throughout the main part of this work.
\bea
I_n ( m_1^2, ...,m_{n-1}^2,m_n^2)& \equiv & \int \frac{d^4 p}{(2\pi)^4} \frac{1}{(p^2 -m_{1}^2)( p^2 -  m_{2}^2)... (p^2 -  m_{n-1}^2)(p^2 - m_n^2)} \nn
\\
&\equiv& \frac{i}{16\pi^2 m_n^{2n - 4}} f_n (x_1,x_2,...,x_{n-1}) \nn
\\
\tilde{I}_n ( m_1^2, ...,m_{n-1}^2,m_n^2)& \equiv&\int \frac{d^4 p}{(2\pi)^4} \frac{p^2}{(p^2 -m_{1}^2)( p^2 -  m_{2}^2)... (p^2 -  m_{n-1}^2)(p^2 - m_n^2)}  \nn
 \\
&\equiv& \frac{i}{16\pi^2 m_n^{2n - 6}} \tilde{f}_n (x_1,x_2,...,x_{n-1}) \nn
\eea
with $x_i \equiv \frac{m_i^2}{m_n^2}$. Here we collect useful relations related to functions $I_{4,5,6}$ and $\tilde{I}_{4,5,6}$:
\bea
\overset{(\sim)}{I}_{\!\!\!4}(M^2,M^2,m^2,m^2)&=&{i\overset{(\sim)}{f}_{\!\!\!4}(x,x,1)\over 16\pi^2m^{4\,(2)}}\nn
\\
\overset{(\sim)}{I}_{\!\!\!5}(M^2,M^2,m^2,m^2,m^2)&=&{i\overset{(\sim)}{f}_{\!\!\!5}(x,x,1,1)\over 16\pi^2m^{6\,(4)}}\nn
\\
\overset{(\sim)}{I}_{\!\!\!6}(M^2,M^2,m^2,m^2,m^2,m^2)&=&{i\overset{(\sim)}{f}_{\!\!\!6}(x,x,1,1,1)\over 16\pi^2m^{8\,(6)}}\nn
\eea
where
\bea
f_4(x,x,1)&=&{2x-2-(x+1)\ln(x)\over (1-x)^3}\nn
\\
\tilde{f}_4(x,x,1)&=&{x^2-1-2x\ln(x)\over (1-x)^3}\nn
\\
f_5(x,x,1,1)&=&{-x^2-4x+5+2(1+2x)\ln(x)\over 2(1-x)^4}\nn
\\
\tilde{f}_5(x,x,1,1)&=&{-5x^2+4x+1+2x(2+x)\ln(x)\over 2(1-x)^4}\nn
\\
f_6(x,x,1,1,1)&=&{-x^3+9x^2+9x-17-6(3x+1)\ln(x)\over 6(1-x)^5}\nn
\\
\tilde{f}_6(x,x,1,1,1)&=&{x^3+9x^2-9x-1-6x(1+x)\ln(x)\over 3(1-x)^5}\nn
\eea
The limits for $x\rightarrow 0$ and $x\rightarrow \infty$ are
\bea
&&\hspace{-0.7cm}\underset{x\rightarrow 0}{\textrm{lim}}f_4(x,x,1)=-\ln(x)-2+\cO(x\ln(x)),\  \underset{x\rightarrow \infty}{\textrm{lim}}f_4(x,x,1)={\ln(x)\over x^2}-{2\over x^2}+\cO(x^{-3}\ln(x))\nn
\\
&&\hspace{-0.7cm}\underset{x\rightarrow 0}{\textrm{lim}}\tilde{f}_4(x,x,1)=-1-2x\ln(x)+\cO(x),\  \underset{x\rightarrow \infty}{\textrm{lim}}\tilde{f}_4(x,x,1)=-{1\over x}+\cO(x^{-2}\ln(x))\nn
\\
&&\hspace{-0.7cm}\underset{x\rightarrow 0}{\textrm{lim}}f_5(x,x,1,1)=\ln(x)+{5\over 2}+\cO(x\ln(x)),\ \underset{x\rightarrow \infty}{\textrm{lim}}f_5(x,x,1,1)=-{1\over 2x^2}+\cO(x^{-3}\ln(x))\nn
\\
&&\hspace{-0.7cm}\underset{x\rightarrow 0}{\textrm{lim}}\tilde{f}_5(x,x,1,1)={1\over 2}+2x\ln(x)+\cO(x),\ \underset{x\rightarrow \infty}{\textrm{lim}}\tilde{f}_5(x,x,1,1)={\ln(x)\over x^2}-{5\over 2x^2}+\cO(x^{-3}\ln(x))\nn
\\
&&\hspace{-0.7cm}\underset{x\rightarrow 0}{\textrm{lim}}f_6(x,x,1,1,1)=-\ln(x)-{17\over 6}+\cO(x\ln(x)),\ \underset{x\rightarrow \infty}{\textrm{lim}}f_6(x,x,1,1,1)={1\over 6x^2}+\cO(x^{-3})\nn
\\
&&\hspace{-0.7cm}\underset{x\rightarrow 0}{\textrm{lim}}\tilde{f}_6(x,x,1,1,1)=-{1\over 3}-2x\ln(x)+\cO(x),\ \underset{x\rightarrow \infty}{\textrm{lim}}\tilde{f}_6(x,x,1,1,1)=-{1\over 3x^2}+\cO(x^{-3}\ln(x))\nn
\eea
and
\bea
f_4(x=1)={1\over 6}\,,&& \tilde{f}_4(x=1)=-{1\over 3}\,,\nn
\\
f_5(x=1)=-{1\over 12}\,,&& \tilde{f}_5(x=1)={1\over 12}\,,\nn
\\
f_6(x=1)={1\over 20}\,,&& \tilde{f}_6(x=1)=-{1\over 30}\nn\,.
\eea

\section{Models of Flavour}
\label{APP:FLAVOURMODELS}

\subsection{Abelian Models}

An inverted hierarchy was invoked some time ago in the literature \cite{pomarol,ckn,cklp} in order to ease the FCNC and CP constraints
in supersymmetric models. To our knowledge, the first class of models 
in which the inverted hierarchy is really predicted are supersymmetric generalisations of abelian flavour models of the Froggatt-Nielsen
type \cite{fn}. These models contain an additional abelian gauge symmetry $U(1)_X$ under which the three fermion generations have different charges (therefore the name horizontal or flavour symmetry), spontaneously broken at a high energy scale by the
vev of (at least) one scalar field $\Phi$, such that $\epsilon = \langle \Phi \rangle / \Lambda \ll 1$ , where $\Lambda$ is the Planck scale or more
generically the scale where Yukawa couplings are generated.  
Quark mass matrices for example, in such models are given, order of magnitude wise, by
\be
h_{ij}^U \ \sim \ \epsilon^{q_i + u_j + h_u} \quad , \quad h_{ij}^D \ \sim \ \epsilon^{q_i + d_j + h_d} \ , \label{abelian1}
\ee
where $q_i$ ($u_i,d_i,h_u,h_d$) denote the $U(1)_X$ charges of the left-handed quarks (right-handed up-quarks, right-handed down-quarks,
$H_u$ and $H_d$, respectively). Quark masses and mixings in the simplest models are given as
\bea
&& \frac{m_u}{m_t} \sim \epsilon^{q_{13}+u_{13}} \quad , \quad \frac{m_c}{m_t} \sim \epsilon^{q_{23}+u_{23}} \quad , \quad \frac{m_d}{m_b} \sim \epsilon^{q_{13}+d_{13}} \quad , \quad \frac{m_s}{m_b} \sim \epsilon^{q_{23}+d_{23}} \ , \nonumber \\
&& \sin \theta_{12} \sim \epsilon^{q_{12}} \quad, \quad \sin \theta_{13} \sim \epsilon^{q_{13}} \quad , \quad 
\sin \theta_{23} \sim \epsilon^{q_{23}} \ . \label{abelian2}
\eea  
A successful fit of the experimental data requires larger charges for the lighter generations
\be
q_1 \ > q_2 \ > q_3 \quad , \quad u_1 \ > u_2 \ > u_3 \quad , \quad d_1 \ > d_2 \ > d_3 \ ,  \label{abelian3}
\ee
one simple example being for example \cite{nir}
\be
q_1 = 3 \ , \ q_2 = 2 \ , \ q_3 = 0 \ , u_1 = 5 \ , \ u_2 = 2 \ , \ u_3 = 0 \ , d_1 = 1 \ , \ d_2 = 0 \ , \ d_3 = 0 \ .  \label{abelian4} 
\ee
Scalar soft masses in abelian flavour models are typically of the form
\be
m_{ij}^2 \ = \ X_i \delta_{ij} \langle D \rangle \ + \ c_{ij} \epsilon^{|q_i-q_j|} ({\tilde m}_F)^2 \ , \label{abelian5} 
\ee 
where $X_i \langle D \rangle$ are D-term contribution for the scalar of charge $X_i$, whereas  the second terms proportional to $({\tilde m}_F)^2$ describe F-term contributions. In the case where D-terms are smaller or at most of the same order
than the F-term contributions, the order or magnitude estimate of the FCNC in the
mass insertion approximation is completely determined by $U(1)$ charges to be
\be
(\delta_{ij}^{u,d})_{LL} \sim \epsilon^{|q_i-q_j|} \ , \
(\delta_{ij}^{d})_{RR} \sim \epsilon^{|d_i-d_j|} \ , \
(\delta_{ij}^{u})_{RR} \sim \epsilon^{|u_i-u_j|} \ . \label{abelian6}   
\ee
If two charges are equal (this is the case for right-handed $d$ quarks above $d_2=d_3$), mass insertion approximation is however not valid anymore.  

\subsection{Non-abelian extension} \label{non-ab}

We present here in some details the model used in Section 4.3.3. The model was proposed in  \cite{Dudas:2013pja} and is a flavour model based on  a $G \times U(1)_{local}$ horizontal symmetry, where $G$ is a discrete nonabelian subgroup of $SU(2)_{global}$. Whereas the discrete nonabelian symmetry is preferable over the continuous $SU(2)_{global}$ for
theoretical reasons, for low-energy flavour physics it was argued in \cite{Dudas:2013pja} that  there is no major difference between the discrete and the continuous case.  

The simplest choice for the flavour charges is to consider an $SU(5)$ invariant pattern $X_{10}$ and $X_{\overline{5}}$, with Higgses uncharged. We need a minimum number of two flavons, an SU(2) doublet $\phi$ with charge $X_\phi$ and an $SU(2)$ singlet $\chi$ with charge $-1$. The total field content is summarised in Table ~\ref{charges}. 
\begin{table}[h]
\centering
\begin{tabular}{c|cccccc|cc}
& $10_a$ & $10_3$ & ${\bar 5}_a$ & ${\bar 5}_3$ & $H_u$ & $H_d$ & $\phi^a$ & $\chi$ \\
\hline
$SU(2)$ & $2$ & $1$ & $2$ & $1$ & $1$ & $1$ & $\overline{2}$  & $1$ \\    
$U(1)$ & $X_{10}$ & $0$ & $X_{\bar 5}$  & $X_3$ &$0$ & $0$ & $X_\phi$ & $-1$    
\end{tabular}
\caption{Flavour group representations of the model.}
\label{charges}
\end{table}
The zero $U(1)$ charge of the 3rd generation ten-plet  takes account of the large top quark Yukawa coupling, whereas $X_3$ is left free, in order to accommodate different values of $\tan\beta$.


The relevant part of the superpotential is given by
\begin{align}
\label{nonab1}
W & = h_{33}^u  H_u  Q_3 U_3+ h_{23}^u Q_a U_3 H_u \frac{\phi^a}{\Lambda} \left(\frac{\chi}{\Lambda} \right)^{X_{10} + X_\phi} + h_{32}^u Q_3 U_a H_u \frac{\phi^a}{\Lambda} \left(\frac{\chi}{\Lambda} \right)^{X_{10} + X_\phi}  \nonumber \\
& + h_{12}^u H_u Q_a U_b \epsilon^{ab} \left(\frac{\chi}{\Lambda} \right)^{2 X_{10} } + h_{22}^u Q_a U_b H_u \frac{\phi^a}{\Lambda} \frac{\phi^b}{\Lambda} \left(\frac{\chi}{\Lambda} \right)^{2 X_{10} + 2 X_\phi} \nonumber \\
&+ h_{33}^d  H_d  Q_3 D_3 \left(\frac{\chi}{\Lambda} \right)^{X_3} + 
h_{23}^d Q_a D_3 H_d \frac{\phi^a}{\Lambda} \left(\frac{\chi}{\Lambda} \right)^{X_{10} + X_3 + X_\phi} + h_{32}^d Q_3 D_a H_d \frac{\phi^a}{\Lambda} \left(\frac{\chi}{\Lambda} \right)^{X_{\bar 5} + X_\phi}  \nonumber \\
&+ h_{12}^d H_d Q_a D_b \epsilon^{ab} \left(\frac{\chi}{\Lambda} \right)^{X_{10}
+ X_{\bar 5} } + h_{22}^d Q_a D_b H_d \frac{\phi^a}{\Lambda} \frac{\phi^b}{\Lambda} \left(\frac{\chi}{\Lambda} \right)^{X_{10} + X_{\bar 5}+ 2 X_\phi} \ . 
\end{align}
 We have imposed here that all exponents are non-negative
 \begin{equation}
 X_{10} \ge 0, X_{3} \ge  0 \ , \ X_{10}+X_{\phi} \ge  0 \ , \ X_{\bar 5} + X_{\phi} \ge  0 \ , \   X_{10} + X_{\bar 5} \ge  0 \ . \label{nonab2}
 \end{equation}
The $h$'s are complex $\mathcal O(1)$ coefficients,  $\Lambda$ is a high flavour scale and  $a,b$ are the $SU(2)$ indices. In the leading order in small parameters, the structure of the K\"ahler potential does not affect the predictions in the fermion sector. Using the flavon vevs
\begin{align}
\langle \phi^a \rangle & = \epsilon_{\phi} \Lambda \begin{pmatrix} 0 \\ 1 \end{pmatrix} 
\ , & \langle \chi \rangle & = \epsilon_{\chi} \Lambda \ , \label{nonab3}
\end{align}
one can calculate masses and mixings in terms of the original parameters.

The Yukawa matrices turn out to be given by
\begin{align}
\label{nonab4}
Y_u & = 
\begin{pmatrix}
0 & h_{12}^u \epsilon_u' & 0 \\
- h_{12}^u  \epsilon_u'  & h_{22}^u \epsilon_u^2& h_{23}^u  \epsilon_u \\
0 & h_{32}^u \epsilon_u & h_{33}^u 
\end{pmatrix} \ , \\
\label{yukawasd}
Y_d & = 
\begin{pmatrix}
0 & h_{12}^d \epsilon_u' \epsilon_d/\epsilon_u & 0 \\
- h_{12}^d \epsilon_u' \epsilon_d/\epsilon_u & h_{22}^d \epsilon_u \epsilon_d & h_{23}^d  \epsilon_3 \epsilon_u \\
0 & h_{32}^d \epsilon_d & h_{33}^d \epsilon_3 
\end{pmatrix} \ , 
\end{align}

with
\begin{align}
\label{nonab5}
\epsilon_u & \equiv \epsilon_{\phi} \epsilon_{\chi}^{X_{10} + X_{\phi}}, & \epsilon_d &  \equiv \epsilon_{\phi} \epsilon_{\chi}^{X_{\bar 5} + X_{\phi}},  &
\epsilon_u' &  \equiv  \epsilon_{\chi}^{2 X_{10}}, & \epsilon_3 &  \equiv  \epsilon_{\chi}^{X_{3}}.
\end{align}

Imposing that the charges are integers then gives a series of possibilities.  A particularly simple possibility, which turns out to be the most successful from the flavour
protection viewpoint is for
\begin{equation}
\label{nonab6}
\epsilon_{\chi} \sim \epsilon_{\phi} \sim 0.02 \ , \qquad X_{10}=X_{\bar 5}=X_3=-X_\phi=1 \ , \qquad \tan \beta = 5\, . 
\end{equation}
The main features of the model are as follows: 

\begin{itemize}
\item The model has $U(1)_X $ D-term contributions which are dominant over the F-term ones
$\langle D \rangle \gg m^2_F$. 
\item  The squark mass matrices are almost diagonal in the flavour basis, with rotation matrices $Z$ which are very close to the identity, compared to the analogous ones for the quarks $U$. In this case, the matrices appearing in the gluino couplings are determined by quark rotations $W \simeq U^{\dagger}$. 
\item Due to the $SU(2)$ original symmetry only broken
by the small parameter $\epsilon_\phi$, the first two generation squarks, both left and
right-handed, are essentially degenerate with mass given by $m_{L1}^2, m^2_{L2} = \langle D 
\rangle$, with non-degeneracies (provided by the flavour breaking) which are negligible. 
\item The main splitting is between the first two and the third generation. For left squarks, there is an hierarchy $m_{L1} \gg m_{3L}$ since the third generation is uncharged under $U(1)_X$
and therefore gets only F-term contributions $m_{3L} \sim m_F$. This is also true for the  right-handed up-type squarks.  
\item The  right-handed down-type squarks are charged and get D-term contributions. In the simplest example we consider here, the third generation is almost degenerate with the   
first two, $m_{3R}^2 =  m_{Rh}^2 + \delta m_{3R}^2 $, where $\delta m_{3R}^2 \sim m_F^2. $
\end{itemize}

The most constraining operator is as usual $Q_4$, from  $\epsilon_K$. For models of the
type described above, the corresponding coefficient in the leading approximation is
given by
\begin{equation}
C_4 = \frac{\alpha_s^2}{3} V_{32}^L {\bar V}_{31}^L V_{32}^R {\bar V}_{31}^R 
\frac{m_{3R}^2-m_1^2}{m_1^4} {\tilde f}_5 ( \frac{m_{3L}^2}{m_1^2} ,
\frac{m_{D}^2}{m_1^2} ) \ . \label{nonab7}
\end{equation}
The relevant rotations are given in the leading approximation by
\begin{eqnarray}
&&V_{32}^L \sim \epsilon_u \quad , \quad {\bar V}_{31}^L \sim \sqrt{\frac{m_d}{m_s}} \epsilon_u
 \ ,  \nonumber \\
&& V_{32}^R \sim \sin \theta_d \quad , \quad {\bar V}_{31}^R \sim  \sqrt{\frac{m_d}{m_s}} \sin \theta_d
\ , \label{nonab8} \end{eqnarray} 
where
\begin{equation}
\tan \theta_d \equiv \frac{|h_{32}^d|\epsilon_d}{|h_{33}^d|\epsilon_3}\   
\label{nonab9}
\end{equation}
is a free parameter of order one fixed to $\tan \theta_d = 0.5$ in order  to correct 
the ratio $V_{ub}/V_{cb}$. 
The product of rotations is therefore given at the leading order in the flavon parameters by
\begin{equation}
V_{32}^L {\bar V}_{31}^L V_{32}^R {\bar V}_{31}^R \sim \frac{m_d}{m_d}
\epsilon_u^2 \sin^2 \theta_d \ , \label{nonab10}
\end{equation}
Notice that the right-handed rotations in (\ref{nonab8}) are large. Because of this lack of suppression, right-handed sbottom has to be heavy. The charge assignment 
(\ref{nonab6}) is then the most advantageous one and realises the minimal
implementation of the natural SUSY spectrum.

\section{B-meson mixing constraints}
\label{APP:Bs}

As discussed in section \ref{SEC:Expressions}, the bounds from $B$-meson mixing can be calculated in the same way as Kaon mixing, using the translations in equation (\ref{EQ:KBDtranslation}) but taking the values in table \ref{BTable}. 
These give us equations of the form 
\begin{align}
C e^{2i\phi} =& 1 + (x + i y) e^{-2i \beta} .
\end{align}
We have limits on $C, \phi$ although they are correlated and it is difficult to use that information. Hence the most conservative bounds that we can set are simply to make sure that $C, \phi$ always lie within their $3\sigma$ ranges. These lead to
\begin{align}
|x_d| <  0.87 \ , \ |y_d| < 0.77 \ , \
|x_s| < 0.3 \ , \ |y_s| < 0.31 \ ,  
\label{EQ:Bmesonbounds}\end{align}
where 
\begin{align}
x_q \equiv {2 \mathrm{Re} \bra B_q^0|\cH_{B_q}|\ov{B}_q^0\ket\over \Delta m_{B_q} (SM) }, \qquad y_q \equiv {2\mathrm{Im} \bra B_q^0|\cH_{B_q}|\ov{B}_q^0\ket\over \Delta m_{B_q} (SM) }
\end{align}
These limits are unlikely to change substantially over the next 20 years: the projected improvement in sensitivity from SuperKEKB with $50 \mathrm{ab}^{-1}$ is from $\pm 0.7 $ to $\pm 0.15$ in $C_{B_d}$ \cite{Aushev:2010bq} (more or less the same as the current UTFIT value), from $\pm 0.1 $ to $\pm 0.03$ in $\phi_{B_d}$ \cite{Aushev:2010bq} (an improvement of about $2$ over the UTFIT present value) while LHCb with $50 \mathrm{fb}^{-1}$ will improve the uncertainty on $\phi_{B_s}$ to $\pm 0.007$ \cite{Bediaga:2012py} -- a factor of $5$ improvement. 

We typically find that the bounds from B-meson mixing are subdominant to those from Kaon mixing; we shall explore this in the mass-insertion approximation and heavy-first-two-generations scenarios below. In this section we shall specialise for clarity to the exactly Dirac gaugino case.

\subsection{Mass insertion approximation}

In the mass insertion approximation, defining
\begin{align}
A \equiv& \left(\frac{\alpha_s}{0.1184}\right)^2 \left( \frac{2000^2\ \mathrm{GeV}^2}{m_{D3}^2} \right) \left(\frac{\tilde{f}_6 (1)}{-1/30} \right)
\end{align}
we find
\begin{align}
x_d + i y_d =& 42 \times A \times \bigg[ 0.27 [\delta_{13}^{LL} \delta_{13}^{LL} + \delta_{13}^{RR} \delta_{13}^{RR}] - 2.1\delta_{13}^{LR}  \delta_{13}^{RL}   -0.13 \delta_{13}^{LL}  \delta_{13}^{RR}  \bigg] \nn\\
x_s + i y_s =& 2 \times A \times  \bigg[ 0.27 [\delta_{23}^{LL} \delta_{23}^{LL} + \delta_{23}^{RR} \delta_{23}^{RR}]- 2.2\delta_{23}^{LR}  \delta_{23}^{RL} -0.13 \delta_{23}^{LL}  \delta_{23}^{RR}  \bigg].
\end{align}
These can be simply translated into bounds using equation (\ref{EQ:Bmesonbounds}). However, if we compare  with the bounds from Kaon mixing we have 
\begin{align}
\frac{\Delta M_K (\mathrm{SUSY})}{\Delta M_K (\mathrm{exp})} =& 280 \times A \times \bigg[ 0.18 [\delta_{12}^{LL} \delta_{12}^{LL} + \delta_{12}^{RR} \delta_{12}^{RR}] - 16\delta_{12}^{LR}  \delta_{12}^{RL}   -4.1 \delta_{12}^{LL}  \delta_{12}^{RR}  \bigg] \nn\\ 
\frac{|\epsilon_K (\mathrm{SUSY})| }{|\epsilon_K (\mathrm{SM})| 0.73} =& 6.7\times 10^4 \times A \times \bigg| \mathrm{Im} \bigg( 0.18 [\delta_{12}^{LL} \delta_{12}^{LL} + \delta_{12}^{RR} \delta_{12}^{RR}] - 16\delta_{12}^{LR}  \delta_{12}^{RL}   -4.1 \delta_{12}^{LL}  \delta_{12}^{RR} \bigg) \bigg|.
\end{align}
We see clearly that the bounds from $\Delta M_K$ and, in particular, $\epsilon_K$ are much more stringent than those from the B meson oscillations.

\subsection{Decoupled first two generations}

We expect that the B-meson mixing bounds should be most relevant in the limit that the first two generations are heavy; here we shall consider that case. For these purposes we can ignore mixing between the first two generations \`a la \cite{gnr} and thus have
\begin{align}
{m_{D3}^2 \over \alpha_s^2}\cL_{dirac} =& {\hat \delta}_{i3}^{LL} {\hat \delta}_{i3}^{LL} {11\over 36}Q_1 + \delta_{i3}^{RR} \delta_{i3}^{RR} {11\over 36} \tilde{Q}_1+ {\hat \delta}_{i3}^{LL}  {\hat \delta}_{i3}^{RR}\Big({5\over 9}Q_5  -{1\over 3}Q_4\Big)
\end{align}
where $i=1$ for $B_d$, $2$ for $B_s$. 

These lead to  (taking the bag factors into account from table \ref{BBagTable})
\begin{align}
x_d + i y_d =& 2260  \times \left(\frac{\alpha_s}{0.1184}\right)^2 \left( \frac{2000^2\ \mathrm{GeV}^2}{m_{D3}^2} \right) \bigg[ 0.27 [{\hat \delta}_{i3}^{LL} {\hat \delta}_{i3}^{LL} + {\hat \delta}_{i3}^{RR} {\hat \delta}_{i3}^{RR}] -0.13 {\hat \delta}_{i3}^{LL}  {\hat \delta}_{i3}^{RR}  \bigg] \nn\\
x_s + i y_s   =& 95 \times \left(\frac{\alpha_s}{0.1184}\right)^2 \left( \frac{2000^2\ \mathrm{GeV}^2}{m_{D3}^2} \right) \bigg[ 0.27 [{\hat \delta}_{i3}^{LL} {\hat \delta}_{i3}^{LL} + {\hat \delta}_{i3}^{RR} {\hat \delta}_{i3}^{RR}] -0.14 {\hat \delta}_{i3}^{LL} {\hat \delta}_{i3}^{RR}  \bigg].
\end{align}
These lead to bounds
\begin{align}
|\mathrm{Re} ( {\hat \delta}_{13}^{LL} {\hat \delta}_{13}^{LL})| < 2.6 \times 10^{-3} \ , \qquad|\mathrm{Im} ( {\hat \delta}_{13}^{LL} {\hat \delta}_{13}^{LL})| < & 2.3 \times 10^{-3} \ , \nn\\
|\mathrm{Re} ( {\hat \delta}_{23}^{LL} {\hat \delta}_{23}^{LL})| < 2.1 \times 10^{-2} \ , \qquad |\mathrm{Im} ( {\hat \delta}_{23}^{LL} {\hat \delta}_{23}^{LL})| < & 2.1 \times 10^{-2} \ . 
\end{align}
Hence the stronger B-meson bounds come from the $B_d$ data rather than $B_s$, but $\epsilon_K$ still provides the strongest constraint on the model parameter space, given in equation (\ref{EQ:decoupledlimits}). These bounds are much weaker than the those from \cite{gnr}, presumably due to the Dirac mass and the factor of $10$ increase in the gaugino mass that we are now required to take. Note that, since there is no square root here, changing the gaugino mass by a factor of ten weakens the bound by a factor of a hundred; whereas in the $\epsilon_K$ case it is only a factor of ten (even for Dirac gauginos). Hence as we make the gauginos heavier we further weaken the relevance of the $B$-mixing compared to $\epsilon_K$.

\section{Input}\label{Input}
\label{APP:INPUT}

Here we collect the Bag factors and B-meson mixing data that we have used in setting bounds. In addition we use 
bag factors and magic numbers given in \cite{Ciuchini:1997bw,Ciuchini:1998ix,utfit} that we have not reproduced here.

\newpage


\begin{table}[!ht]
\begin{minipage}[b]{0.55\textwidth}
\begin{center}
\begin{tabular}{|c|c|c|}\hline
Parameter & Value & Ref.\\\hline
$\alpha_s(M_Z)$ & 0.1184 &\cite{pdg}\\\hline
$f_K$ & $0.1598\,\mathrm{GeV}$ & \cite{pdg}\\\hline
$m_K$ & 0.497672 GeV & \cite{pdg}\\\hline
$m_s(2\,\textrm{GeV})$ & $0.095\,\mathrm{GeV}$ & \cite{pdg}\\\hline
$m_d(2\,\textrm{GeV})$ & $0.0048\,\mathrm{GeV}$ & \cite{pdg}\\\hline
$\Delta m_K^{\mathrm{exp}}$ & $(3.484 \pm 0.006)\times 10^{-15}\,\mathrm{GeV}$ &\cite{pdg} \\\hline
$|\epsilon_K^\mathrm{exp}|$ & $(2.228 \pm 0.011) \times 10^{-3}$ &\cite{pdg} \\\hline
$|\epsilon_K^{\textrm{SM}}|$ & $(2.04 \pm 0.19) \times 10^{-3}$ & \cite{utfit}\\\hline
\end{tabular}
\end{center}
\caption{Input used for Kaon bounds}
\label{KInput}
\end{minipage}
\begin{minipage}[b]{0.45\textwidth}
\begin{center}
\begin{tabular}{|c|c|}\hline
Parameter & Value\\\hline
 $B_1$ & 0.60 \\\hline
 $B_2$ & 0.66 \\\hline
 $B_3$ & 1.05 \\\hline
 $B_4$ & 1.03 \\\hline
 $B_5$ & 0.73 \\\hline
 \end{tabular}
\end{center}
\caption{Bag numbers for Kaons \cite{Allton:1998sm}.}
\label{KBagInput}
\end{minipage}
\end{table}

\begin{table}[!hc]
\begin{minipage}[b]{0.5\textwidth}
\begin{center}
\begin{tabular}{|c|c|}\hline
Parameter & Value\\\hline
$\Delta m_{B_d}$ (SM) & $3.36 \pm 0.03 \times 10^{-13}$ GeV \\\hline
$\beta_{B_d}$ (SM)& $0.426\pm 0.031$\\\hline
$\Delta m_{B_s}$ (SM)& $117 \pm 0.16 \times 10^{-13}$ GeV\\\hline
$\beta_{B_s}$ (SM)& $0.0187 \pm 0.0007$\\ \hline
$C_{B_d}$ &  $1.07 \pm 0.17 $\\\hline
$\phi_{B_d}$ & $-0.035 \pm 0.056$\\\hline
$C_{B_s}$ & $1.066 \pm 0.083 $\\\hline
$\phi_{B_s}$ & $0.010 \pm 0.035$\\\hline
$m_{B_d}$ & $5279.58 \pm 0.17$ MeV\\\hline
$m_{B_s}$ & $5366.77 \pm 0.24$ MeV\\\hline
$m_b$ & $4.18 \pm 0.03 $ GeV ($\ov{MS}$)\\\hline
$m_s$ & $95 \pm 5 $ MeV\\\hline
$m_d$ & $4.8^{+0.5}_{-0.3} $ MeV\\\hline
$f_{B_d}$ & $186 \pm 4$ MeV\\\hline
$f_{B_s}$ & $224 \pm 5 $ MeV\\\hline
$\left( \frac{m_{B_d}}{m_b+m_d}\right)^2$ & $1.59$\\\hline
$\left( \frac{m_{B_s}}{m_b+m_d}\right)^2$ & $1.64$ \\\hline
\end{tabular}
\end{center}
\caption{Data for B-meson bounds, taken from UTFIT website \cite{utfit} and \cite{Dowdall:2013tga} (for the meson decay constants).}
\label{BTable}
\end{minipage}
\begin{minipage}[b]{0.05\textwidth}
\hspace{0.05\textwidth}
\end{minipage}
\begin{minipage}[b]{0.4\textwidth}
\begin{center}
\begin{tabular}{|c|c|}\hline
Parameter & Value\\\hline
$B_1^d$ & $0.87(4)$ \\
$B_2^d$ & $0.79(2)$ \\
$B_3^d$ & $0.92(6)$ \\
$B_4^d$ & $1.15(3)$ \\
$B_5^d$ & $1.72(4)$ \\\hline
$B_1^s$ & $0.87(2)$ \\
$B_2^s$ & $0.80(1)$ \\
$B_3^s$ & $0.93(3)$ \\
$B_4^s$ & $1.16(2)$ \\
$B_5^s$ & $1.75(3)$ \\\hline
\end{tabular}
\end{center}
\vspace{\baselineskip}
\caption{Bag numbers for B mesons from \cite{Becirevic:2001yv}.}
\label{BBagTable}
\end{minipage}
\end{table}


\end{document}